\documentclass[12pt]{article}
\usepackage{amsmath,amsbsy,amssymb, amsfonts, amsthm, latexsym, graphics, graphicx, tabularx}
\usepackage{multirow,multicol,subfigure,booktabs,color}
\usepackage{psfrag,epsfig}
\usepackage{natbib}
\usepackage{mathtools}
\usepackage{color,hyperref}
\usepackage{comment}
\usepackage{bigints}
\definecolor{darkblue}{rgb}{0.0,0.0,0.55}
\hypersetup{colorlinks,breaklinks,	
	linkcolor=darkblue,urlcolor=darkblue,
	anchorcolor=darkblue,citecolor=darkblue}
 \usepackage[table,xcdraw]{xcolor}
\usepackage[margin=10pt, font=small,labelfont=bf,labelsep=period]{caption}

\usepackage{lastpage}
\usepackage{fancyhdr}
\newcommand{\ben}{\begin{enumerate}}
	
	\newcommand{\een}{\end{enumerate}}

\newcommand{\beq}{\begin{equation}}

\newcommand{\eeq}{\end{equation}}

\newcommand{\beqn}{\begin{eqnarray}}

\newcommand{\eeqn}{\end{eqnarray}}

\newcommand{\bea}{\begin{eqnarray*}}
	
	\newcommand{\eea}{\end{eqnarray*}}

\newcommand{\bs}{\boldsymbol}

\newcommand{\argmax}{\operatornamewithlimits{argmax}}

\usepackage{dsfont}

\definecolor{labelkey}{rgb}{0.6,0,1}

\usepackage[ruled]{algorithm2e}

\usepackage{algorithmic}

\begin{document}
	\begin{center}
		\Large{\textbf{Detecting differentially methylated regions in bisulfite sequencing data using quasi-binomial mixed models with smooth covariate effect estimates}}
	\end{center}
	\begin{center}	
{
	Kaiqiong Zhao$^{1, 2}$, Karim Oualkacha$^{3}$, Lajmi Lakhal-Chaieb$^{4}$,   Aur\'elie Labbe$^5$, \\ Kathleen Klein$^{2}$, Sasha Bernatsky$^{6,7}$, Marie Hudson$^{2,6}$, In\'es Colmegna$^{6,7}$, Celia M.T. Greenwood$^{1, 2, 8, 9}$    
  	\\
  	$^1$Department of Epidemiology, Biostatistics and Occupational Health, \\ McGill University \\   
  	$^2$Lady Davis Institute for Medical Research, Jewish General Hospital\\ 
  	$^{3}$D\'epartement de Math\'ematiques,  Universit\'e du Qu\'ebec \`a Montr\`eal\\
  	$^4$D\'epartement de Math\'ematiques et de Statistique,  Universit\'e Laval \\
  	$^5$D\'epartement des Sciences de la D\'ecision, HEC Montr\`eal \\ 
  	$^6$Department of Medicine, McGill University\\
  	$^7$The Research Institute of the McGill University Health Centre\\ 
  	$^{8}$Department of Human Genetics,  McGill University \\ 
  	$^{9}$Gerald Bronfman Department of Oncology,  McGill University\\
  }
  	\end{center}
  	\begin{center}
January 18, 2021
	\end{center}
	
	\normalsize
\begin{abstract}
\noindent
Identifying disease-associated changes in DNA methylation can help to gain a better understanding of disease etiology. Bisulfite sequencing technology allows the generation of methylation profiles at single base of DNA. We previously developed a method for estimating smooth covariate effects and identifying differentially methylated regions (DMRs) from bisulfite sequencing data, which copes with experimental errors and variable read depths; this method utilizes the binomial distribution to characterize the variability in the methylated counts. However, bisulfite sequencing data frequently include low-count integers and can exhibit over or under dispersion relative to the binomial distribution. We present a substantial improvement to our previous work by proposing a quasi-likelihood-based regional testing approach which accounts for multiplicative and additive sources of dispersion. We demonstrate the theoretical properties of the resulting tests, as well as their marginal and conditional interpretations. Simulations show that the proposed method provides correct inference for smooth covariate effects and captures the major methylation patterns with excellent power.
\end{abstract}
	\section{Introduction}
	Conceptually, the emergence of a disease phenotype is believed to stem from the combined effects of genetic predisposition and environmental exposures \citep{ober2011gene}. A plausible mechanism behind this gene-environment interplay is epigenetic modification, which regulates gene activity through modifications of DNA accessibility.  Epigenetics may explain how exposures leave heritable marks on the genome that impact disease susceptibility \citep{jaenisch2003epigenetic}. Therefore, increased understanding of epigenetic-disease association could lead to novel insights into disease causation and possible therapies \citep{feinberg2007phenotypic}.
	% UPdate the reference, Recently, epigenetic changes have been shown to associate with, epigenetic clockes, for example
	
	The most studied epigenetic mark is DNA methylation, which involves the covalent addition of a methyl group to a cytosine nucleotide.  DNA methylation, in the mammalian genomes, occurs predominantly at cytosine-guanine dinucleotides (i.e. CpG sites) \citep{lister2009human}. Methylation of CpG-rich promoters can silence gene expression by preventing transcriptional factor binding to DNA \citep{choy2010genome}. More generally, DNA methylation has the potential to activate or repress gene expression, depending on whether the mark inactivates a positive or negative regulatory element \citep{jones1999dna}.  Known or suspected drivers behind methylation alterations include genetic variations \citep{mcrae2014contribution}, environmental toxins \citep{hanson2008developmental}, external stressors \citep{dolinoy2007maternal}  and aging \citep{horvath2013dna}. There is also evidence that localized abnormal methylation is strongly linked to many diseases, including breast cancer \citep{hu2005distinct}, autism spectrum disorder \citep{dunaway2016cumulative}, and systemic autoimmune disease \citep{kato2005genetic}.
	% UPDATE the reference
	
	High-resolution, large-scale measurement of DNA methylation is now possible with recent advances in bisulfite sequencing (BS-seq) protocol, which is implemented either genome-wide or in targeted regions.  Although whole-genome bisulfite sequencing (WGBS) allows a comprehensive characterization of the methylation landscape, it is inefficient for large-scale studies as only 20\% or less of CpGs are thought to have variable methylation across individuals or tissues \citep{ziller2013charting}. On the other hand, Targeted Custom Capture Bisulfite Sequencing (TCCBS) platform enables a comprehensive yet cost-effective interrogation of functional CpGs in disease-targeted tissues or cells \citep{allum2015characterization}. This approach has been successfully used to identify novel disease-associated epigenetic variants \citep{shao2019rheumatoid, allum2019dissecting, ziller2016targeted}. In this work, we aim to improve sensitivity to detect, among all the regions targeted by TCCBS, differentially methylated regions (DMRs) that are associated with phenotypes or traits.
    
Like other sequencing experiments, the raw data from TCCBS are short sequence reads. After proper alignment and data processing, the methylation level at a single cytosine can be summarized as a pair of counts: the number of methylated reads and the total number of reads covering the site, i.e. read depth. Such data possess several challenges for statistical analysis. Typically, read depth varies drastically across sites and individuals, which leads to measures with wide-ranging precision and many missing values \citep{sims2014sequencing}. Additional statistical challenges are created by the strong spatial correlations observed in methylation levels at neighboring CpG sites \citep{hansen2012bsmooth, rackham2017bayesian, korthauer2017detection, shokoohi2018hidden}, as well as the possibility of data errors, arising from excessive or insufficient bisulfite treatment or other aspects of the sequencing processes \citep{cheng2013classification,lakhal2017smoothed}. Furthermore, in addition to the trait of interest (e.g. disease or treatment group), other factors, such as age \citep{horvath2013dna}, batch effects \citep{leek2010tackling}, or cell-type mixture proportions (for mixed tissue samples) \citep{mcgregor2016evaluation} have effects on methylation levels. Hence, it is desirable to adjust methylation signals for multiple covariates simultaneously. 
 
    \begin{figure}[h!]
    	\centering 
    	(A) \\
    	\includegraphics[width=0.8\linewidth]{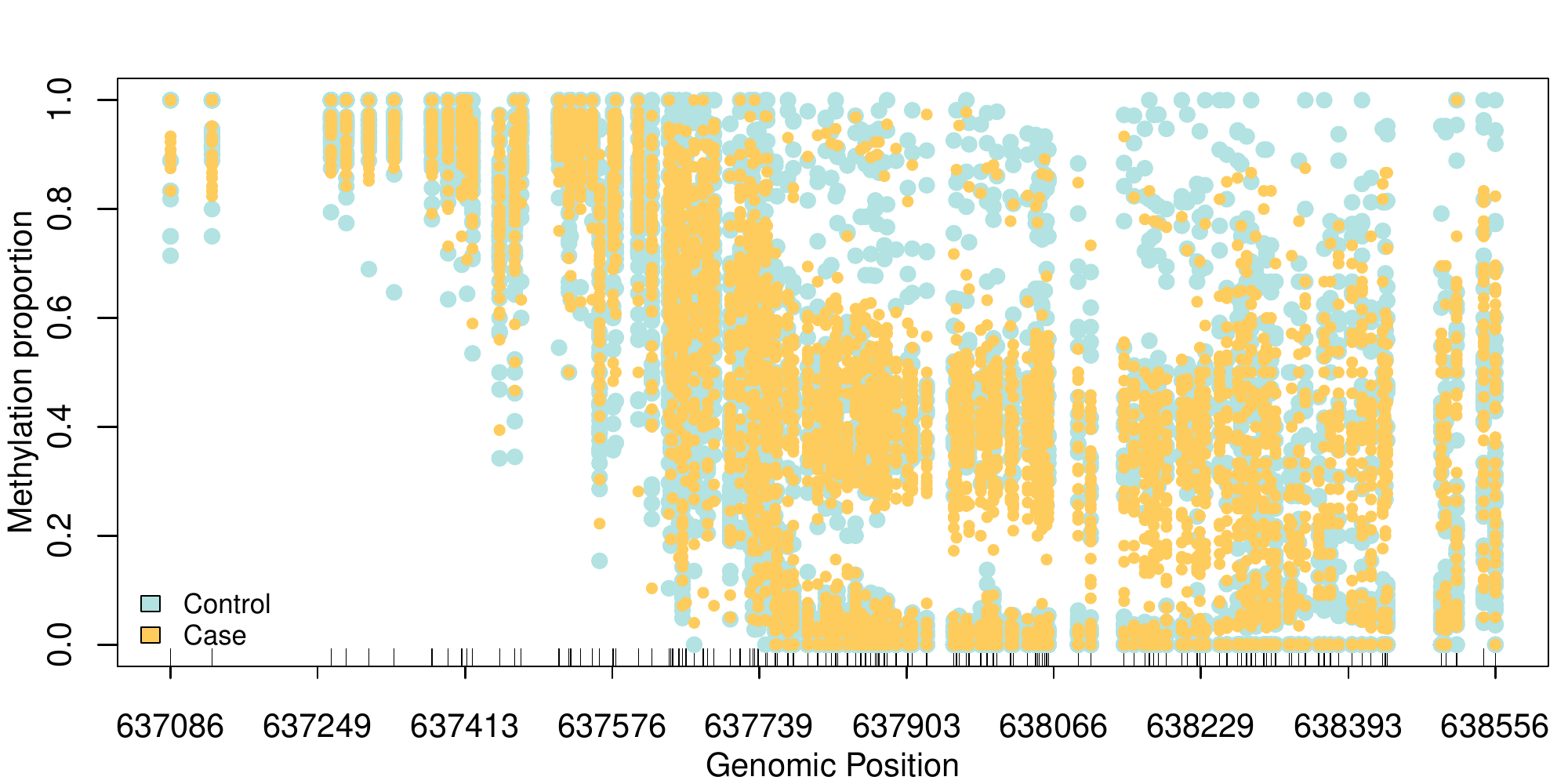}\\
    	\centering
    	(B) \qquad  \qquad \qquad  \qquad \qquad  \qquad (C)\\
    	\includegraphics[width=0.8\linewidth]{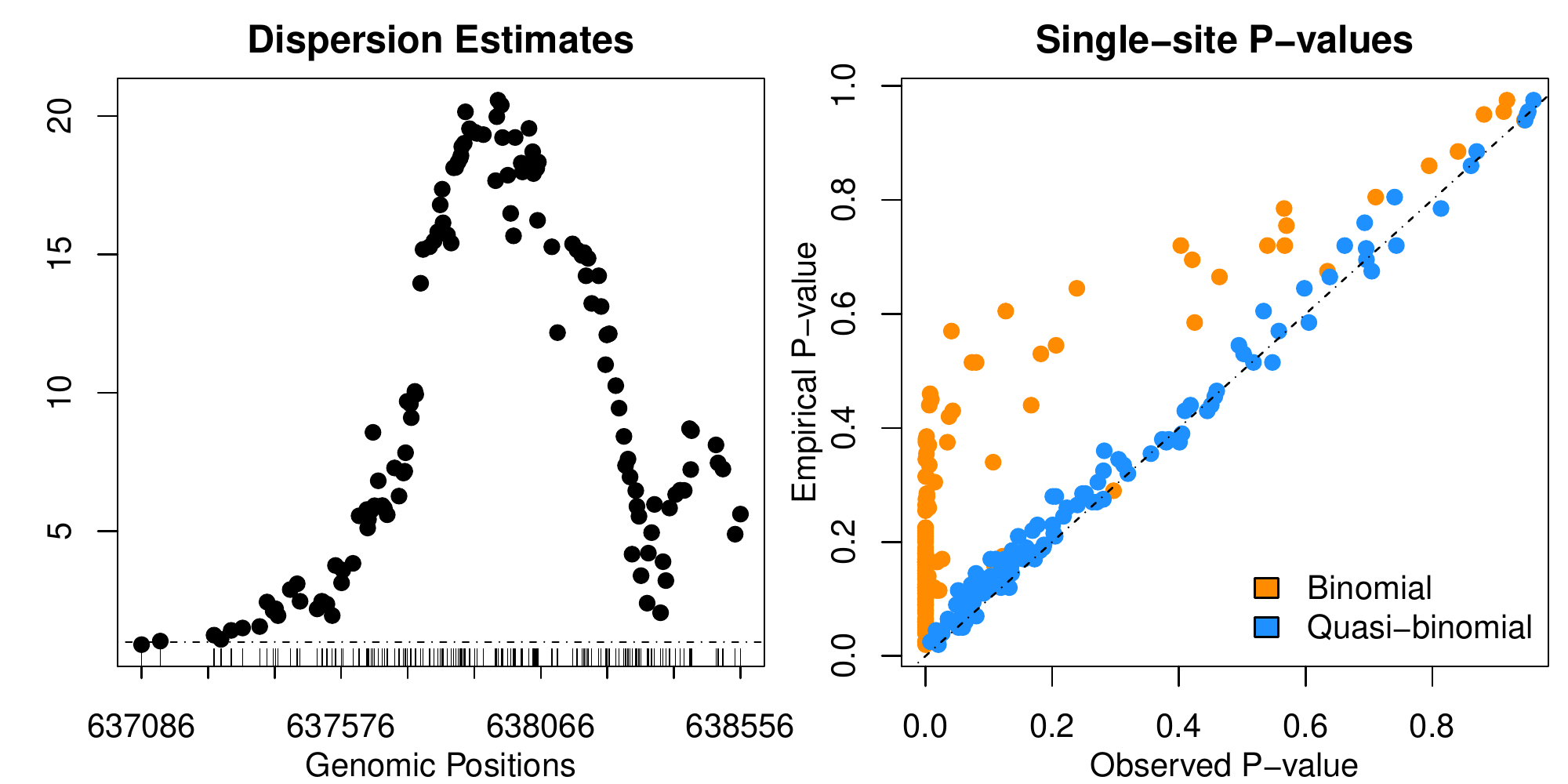}
    	\caption{\textit{Illustration of observed dispersion in a targeted region that underwent bisulfite sequencing.} (A) Observed methylation proportions in one region for two groups of samples (yellow and blue); data are fully described in Section \ref{data}. (B) Estimated dispersion for each CpG site from a single-site quasi-binomial GLM. (C) Single-site p-values for methylation difference between the two groups. Horizontal axis are the p-values estimated from either binomial (ignoring dispersion) or quasi-binomial (accounting for dispersion) GLMs. Vertical axis shows the empirical p-values computed from 199 permutations; the empirical p-value is a benchmark for valid statistical tests. (Single-site beta-binomial regression models generate similar dispersion estimate pattern and p-value distribution to quasi-binomial GLM).}
    	\label{fig:fig1-2-univariatedispersionobsemppvals}
    \end{figure}
     %at 129 CpGs
     % (1470bp, 129 CpG sites)
     
To detect truly differentially methylated regions without finding false associations, it is crucial to accurately account for the sources of variability across individuals. We ran into this issue in a recent analysis of methylation profiles and anti-citrullated protein antibodies (ACPA). Figure \ref{fig:fig1-2-univariatedispersionobsemppvals} (A) illustrates methylation proportions in a targeted region for samples from this study. (A full description of the study, referred to as the ACPA dataset, is in Section \ref{data}). Clearly, dispersion is much larger between samples in the blue group. In panel (C), it can be seen that p-values testing for methylation differences, assuming a binomial mean-variance relationship are much too small. In contrast, allowing for dispersion through a quasi-binomial model provides p-values in line with null expectation for this region. As such, the restrictive mean-variance relationship implied by a binomial generalized linear model (GLM) may not adequately accommodate the data variability, and thus can lead to inflation of false positives. This is known as over or underdispersion, i.e. data presenting greater or lower variability than assumed by a GLM model. 
% SEE method section
% UPON upon modeling data with biological replicates within groups,
 
%Additionally, detection of differential methylation requires adequate accounting for the biological variance among samples of similar characteristics \citep{hansen2011sequencing}. For example, in simple two-group comparisons, a large observed methylation difference for a site varying less across within-group individuals is likely more important than the same difference for a highly variable site. 

Moving in this direction, we have developed a SmOoth ModeliNg of BisUlfite Sequencing (SOMNiBUS) method to detect DMRs in targeted bisulfite sequencing data \citep{zhao2020novel}. The method provides a general framework of analysis, and simultaneously addresses regional testing, estimation of multiple covariate effects, adjustment for read depth variability and experimental errors. Specifically, \cite{zhao2020novel} proposed a hierarchical binomial regression model, which allows covariate effects to vary smoothly along genomic position. A salient feature of SOMNiBUS is its one-stage nature.  Several existing methods first smooth methylation data and then, in a second stage, estimate covariate effects based on the smoothed data \citep{hansen2012bsmooth, lakhal2017smoothed,hebestreit2013detection}, and this two-stage framework could lead to biased uncertainty estimates. In contrast, SOMNiBUS collapses smoothing and testing steps into a single step, and achieves accurate statistical uncertainty assessment of DMRs.  That said, its underlying binomial assumption may be overly restrictive and is only applicable when data exhibit variability levels that are similar to those anticipated based on a binomial distribution
%data exhibit a similar amount of variability to the binomial distribution 
(such as data from inbred animal or cell line experiments). In this work, we propose an extension of SOMNiBUS, which maintains all the good properties of the standard SOMNiBUS, and at the same time explicitly allows the variability in regional methylation counts to exceed or fall short of what binomial model permits. 

%YOU need to take a sentence or two here talking about your theoretical developments to "sell" the novelty here

% regression residuals
The importance of accounting for dispersion in BS-seq data has been well recognized in analysis of single CpG sites. Faced with dispersion in discrete data analysis, one commonly used option is to convert the methylated and total counts to proportions. In this way, testing of differentially methylated single CpG sites can be done via the two sample t-test \citep{hansen2012bsmooth} or beta regression \citep{hebestreit2013detection}, both of which allow direct computation of (within-group) sample variation. However, this conversion loses information, since it fails to distinguish between noisy and accurate measurements \citep{wu2015detection}, often as a consequence of the stochasticity of read depth, and also disregards the discrete nature of the data \citep{lea2015flexible}. On the other hand, there are approaches for DNA methylation analysis that directly model counts while accounting for dispersion.  These count-based approaches use either \textit{additive} overdispersion models, or \textit{multiplicative} under- or overdispersion models to describe the variation driving the dispersion \citep{browne2005variance}. In a multiplicative model, one includes a multiplicative scale factor, i.e. the dispersion parameter, in the variance of the binomial response. Thus, the dispersion inflates or deflates the variance estimates of the covariate effect by the multiplicative factor. Such approaches include the quasi-binomial regression model \citep{akalin2012methylkit} and the beta-binomial regression model \citep{dolzhenko2014using, feng2014bayesian, park2014methylsig, park2016differential}. In contrast, additive overdispersion methods add a subject-level random effect (RE) to capture the extra-binomial variation among individual observations.  Both ABBA \citep{rackham2017bayesian} and MACAU \citep{lea2015flexible}, that use binomial mixed effect models fall in this category. An advantage of the multiplicative approach, particularly the quasi-binomial model, is that it naturally allows for both overdispersion and underdispersion, whereas the additive model only allows overdispersion. On the other hand, the additive overdispersion approach links directly with a multilevel model and can be readily extended to analyze data with a hierarchical or clustering structure.

% \cite{harrison2015comparison} showed that the efficacy of the dispersion-correction approach depends on the process that generates the dispersion; they found that random effect model failed to cope with the overdispersion generated from a beta-binomial mixture model and beta-binomial model tended to underestimate effect sizes when modelling non-beta-binomial data. 

\begin{figure}[h!]
	\centering
	(A) \qquad  \qquad \qquad  \qquad \qquad  \qquad  (B)
	\includegraphics[width=0.8\linewidth]{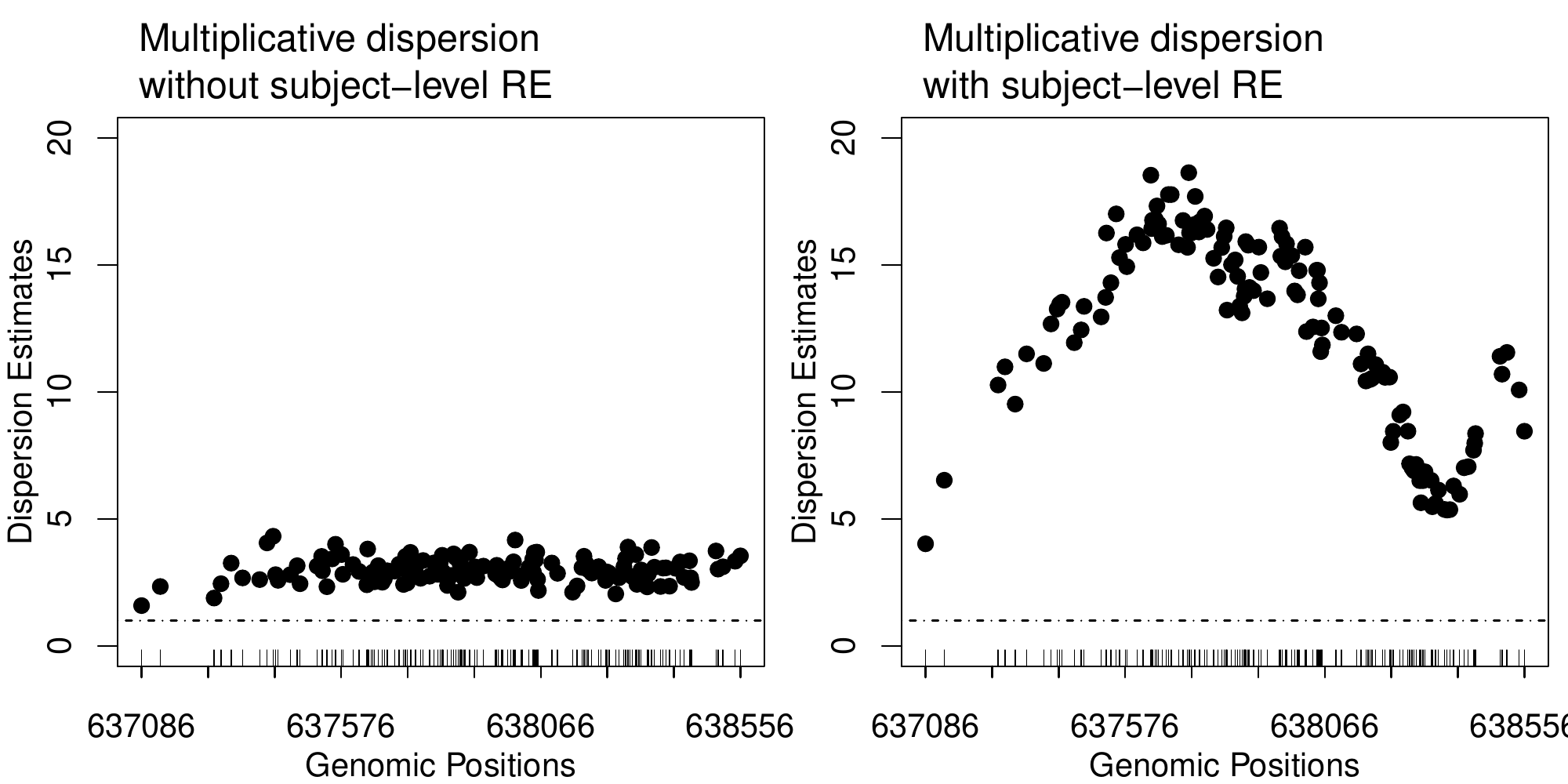}
	\caption{\textit{A byproduct of introducing a subject-level RE, on top of a multiplicative dispersion parameter, to a model with smooth covariate effects is a regional dispersion pattern of varying degree.} Estimated dispersion for each CpG site obtained from a single-site quasi-binomial GLM, for two simulated regional methylation datasets: (A) data were simulated from a multiplicative-dispersion-only model ($\phi = 3, \sigma_0^2 =0$), and (B) data were simulated from a model with both a multiplicative dispersion and a subject-level RE ($\phi = 3, \sigma_0^2 =3$); see Section \ref{model-section} for detailed model formulations and notation definitions.}
	\label{fig:fig2-dispersionpatternbeta1actual}
\end{figure}

The challenge of accounting for dispersion when detecting DMRs is further complicated by several factors. Firstly, even within a small genomic region, different CpG sites may exhibit different levels of dispersion and strong spatial correlation (Figure \ref{fig:fig1-2-univariatedispersionobsemppvals} B). Hence, a multiplicative dispersion model with a common dispersion parameter does not adequately capture the dispersion heterogeneity across loci (Figure \ref{fig:fig2-dispersionpatternbeta1actual} A). In addition, challenges are presented by the complex correlation structure in the regional methylation data. Apart from the spatial correlations among neighboring CpGs, there are additional correlations among methylation measurements on the same subject. Ignoring this within-subject correlation could lead to overestimation of precision and invalid statistical tests \citep{cui2016if}. One means to accommodate such a correlation structure is to add a subject-level RE that can also capture the overdispersion induced by independent variation across different subjects. Furthermore, when modeling discrete data with a hierarchical structure, extra non-structural specific random dispersion can arise, beyond that introduced by the subject-level RE \citep{breslow1993approximate, molenberghs2007extended, vahabi2019joint}, and thus, often, parametric distributions with restrictive mean-variance relations poorly describe the outcomes for individual subjects (i.e. the conditional distribution of outcome given the RE) \citep{molenberghs2010family, molenberghs2012combined,  ivanova2014model}. Hence, properly addressing both multiplicative and additive sources of dispersion in methylation data is essential for making reliable inference at the region level.

\begin{table}[h!]
	\caption{List of existing DNA methylation analytical methods and our proposal with their capabilities.}
	\label{methodsreview}
	\scalebox{0.7}{
	\begin{tabular}{lccccccccc}
		\hline
		\rowcolor[HTML]{C0C0C0} 
		Method  & regional &\begin{tabular}[c]{@{}l@{}}one-\\ stage\end{tabular} &\begin{tabular}[c]{@{}l@{}}count-\\ based\end{tabular}& \begin{tabular}[c]{@{}l@{}}read-depth \\ variability\end{tabular} & 
		\begin{tabular}[c]{@{}l@{}} adjust for \\ confounding\end{tabular}&
	 \begin{tabular}[c]{@{}l@{}}within-subject\\ correlation\end{tabular} & \begin{tabular}[c]{@{}l@{}}non-structural\\ dispersion\end{tabular}  & 
	 	\begin{tabular}[c]{@{}l@{}}varying levels\\ of dispersion\\ across loci\end{tabular} &
		\begin{tabular}[c]{@{}l@{}}experimental\\ errors\end{tabular} \\
		dSOMNiBUS  &  
		$\checkmark$   & 
		$\checkmark$&
		  $\checkmark$   &  
		  $\checkmark$  &           $\checkmark$                                                       &        $\checkmark$                                                                                  &                                            $\checkmark$                           &                 $\checkmark$       &$\checkmark$                                                 \\
		\rowcolor[HTML]{EFEFEF} 
			SOMNiBUS & 		
			$\checkmark$& 
				$\checkmark$	&         
			$\checkmark$&                                           
			$\checkmark$& 
			$\checkmark$&                                               
			&               
			&                   
			&   $\checkmark$ \\  
		BSmooth    &
		 $\checkmark$  &
		 &
		              &
		  $\checkmark\kern-1.1ex\raisebox{.7ex}{\rotatebox[origin=c]{125}{--}}$&
		  &
		 &                    
		 $\checkmark$ &          
		  $\checkmark$ &    \\
		\rowcolor[HTML]{EFEFEF}                                                       
		SMSC       &  
		 $\checkmark$ & 
		 &         
		 &                                           
		 $\checkmark\kern-1.1ex\raisebox{.7ex}{\rotatebox[origin=c]{125}{--}}$&  
		 &                                              
		 &               
	 $\checkmark$	&                  
		$\checkmark$ &  $\checkmark$      \\
		dmrseq     &    	
		$\checkmark$ &          
		$\checkmark$&    
		&                                       
		$\checkmark$ &   
		$\checkmark$&                                             
		$\checkmark$ &               
		$\checkmark$ &                  
	 &       \\
	\rowcolor[HTML]{EFEFEF} 
		Biseq      &    		 
		$\checkmark$ &  
        &        
	&                                  
		$\checkmark\kern-1.1ex\raisebox{.7ex}{\rotatebox[origin=c]{125}{--}}$&  
		$\checkmark$&                                              
		$\checkmark$&               
		$\checkmark$&                  
		$\checkmark$&       \\
		GlobalTest & 		
		$\checkmark$& 
			$\checkmark$&         
		&                                           
		&                
		$\checkmark$&                                
	NA$^\dagger$&               
	NA$^\dagger$&                   
	NA$^\dagger$&    \\
\rowcolor[HTML]{EFEFEF} 
		ABBA & 		
		$\checkmark$&          
		$\checkmark$&
		$\checkmark$&                                           
		$\checkmark$&   
		&   
			$\checkmark$&                                            
		$\checkmark$&                                
		$\checkmark$&    \\
		MACAU & 		
		&  
		$\checkmark$&        
		$\checkmark$&                                           
		$\checkmark$&  
		$\checkmark$	&   
				NA$^\ddagger$&                                              
		$\checkmark$&                               
		$\checkmark$&  \\
		     \hline      
		     \multicolumn{10}{l}{	$\checkmark\kern-1.1ex\raisebox{.7ex}{\rotatebox[origin=c]{125}{--}}$: These three methods are of a two-stage nature. Their smoothing stage indeed accounts for read-depth variability,  but their} \\  
		   \multicolumn{10}{l}{\;\;\;\;\; testing stage, which relies on t-test or beta regression, ignores the read-depth variability.}\\
	\multicolumn{10}{l}{$\dagger:$ GlobalTest treats methylation levels at multiple loci as covariates and trait of interest as outcome. It is not necessary for} \\
	\multicolumn{10}{l}{\;\;\;\; GlobalTest to account for  the three features on covariance structure of methylation across samples and loci.}   \\
		\multicolumn{10}{l}{$\ddagger:$ MACAU is a single-site method and within-subject correlation is irrelevant when analyzing individual sites} \\
		\multicolumn{10}{l}{\;\;\;\;\; one at a time. }                                                  
	\end{tabular}
}
\end{table}

Given our preliminary exploration of dispersion in the ACPA dataset, we recognized the need for a regional one-stage method of analysis that accommodates both the hierarchically-induced overdispersion (and/or correlation) and the extra unstructured individual dispersion. This desired method should also simultaneously address discrete nature of the data, varying strength of dispersion across a region, estimation of multiple covariate effects, adjustment for read depth variability and experimental errors. However, to the best of our knowledge, none of the existing methods meet all aforementioned objectives (Table \ref{methodsreview}). For example, dmrseq \citep{korthauer2017detection}, which fits a generalized least squares regression model with autoregressive error structure to the transformed methylation proportions, accommodates both within-subject correlation and non-structural dispersion, but it assumes a constant dispersion parameter for all loci in a region. Biseq \citep{hebestreit2013detection} is capable of capturing the covariance structure of regional methylation data (by estimating the variogram of site-specific test statistics). However, this method separates smoothing and inference steps and its final significance assessment does not account for the uncertainty in the smoothing step.

To overcome the limitations and challenges of existing methods, we propose a novel approach for identifying DMRs, dSOMNiBUS (dispersion-adjusted SmOoth ModeliNg of BisUlfite Sequencing). Our strategy directly models raw read counts while accounting for all (known) sources of data variability and varying degree of dispersion across loci, thus providing accurate assessments of regional statistical significance.

Specifically, we propose a quasi-binomial mixed model to describe bisulfite sequencing data, which allows covariate effects to vary smoothly along genomic positions, and specially, captures the extra-binomial variation by the \textit{combination} of a subject-specific RE (i.e an additive overdispersion) and a multiplicative dispersion. The RE term accounts for between-sample heterogeneity, and at the same time enables flexible dispersion patterns in a region (Figure \ref{fig:fig2-dispersionpatternbeta1actual} B), which is highly plausible in methylation data (Figure \ref{fig:fig1-2-univariatedispersionobsemppvals} B).
%Introducing separate random effects to each subject can accommodate the association between methylation measurements on the same subject and at the same time  flexible dispersion patterns that vary along genomic regions (Figure \ref{fig:fig2-dispersionpatternbeta1actual} (B)), and thus adjusting routinely to real-world dispersion scenarios (Figure \ref{fig:fig1-2-univariatedispersionobsemppvals} (B)). 
The multiplicative dispersion, on the other hand, explicitly allows the variability in individual subject's methylation levels to exceed or fall short of what binomial distribution assumes, and thus captures the extra dispersion that cannot explained by RE. In addition, our approach accounts for possible data errors in the observed methylated counts. Specifically, we assume that the observed read counts arise from an unobserved latent true methylation state compounded by errors. We then build a specialized expectation-maximization (EM) algorithm for the quasi-binomial mixed model to make inference about DMRs in the presence of data errors. 

%We present a substantial improvement to our previous work by proposing a quasi-likelihood-based regional testing approach which accounts for multiplicative and additive sources of dispersion. We demonstrate the theoretical properties of the resulting tests, as well as their marginal and conditional interpretations. Simulations show that the proposed method provides correct inference for smooth covariate effects and captures the major methylation patterns with excellent power.
%AGAIN here, I think you should sell your work a bit more and talk about theoretical developments and/or how difficult this is to do. Mention the marginal/conditional developments, use of the Laplace approximation, the region based tests and maybe the treatment of errors with dispersion

%We compared dSOMNiBUS’s performance with existing methods using both real bisulfite sequencing data sets and simulated datasets under various settings. 

\section{Results}
  \subsection{The smoothed quasi-binomial mixed model}
   \label{model-section}
 Here we present our model for describing regional methylation data. Details on the algorithm, and  the inference method for the model, are provided in Section \ref{method_detail}.
 
We consider  DNA methylation measures over a targeted genomic region from $N$ independent samples. 
  Let $m_i$ be the number of CpG sites for the $i^{th}$ sample, $i = 1, 2, \ldots N$. We write $t_{ij}$ for the genomic position (in base pairs) for the $i^{th}$ sample at the $j^{th}$ CpG site, $j = 1, 2, \ldots, m_i$. 
  Methylation levels at a site are quantified by the number of methylated reads and the total number of reads. 
  We define $X_{ij}$ as the total number of reads aligned to CpG $j$ from sample $i$. 
  We denote the \textit{true} methylation status for the $k^{th}$ read obtained at CpG $j$ of sample $i$ as $S_{ijk}$, where $k = 1, 2, \ldots X_{ij}$. For a single DNA strand read, $S_{ijk}$ is binary and we define  $S_{ijk} = 1$ if the corresponding read is methylated and $S_{ijk} = 0$ otherwise.
We additionally denote the \textit{true} methylated counts at CpG $j$ for sample $i$ with $S_{ij} = \sum_{k=1}^{X_{ij}} S_{ijk}$, summing over all reads aligned to position $t_{ij}$.
  Furthermore, we assume that we have the information on $P$ covariates for the $N$ samples, denoted as $\bs{Z_i} = \left(Z_{1i}, Z_{2i}, \ldots Z_{Pi} \right)$, for $i = 1, 2, \ldots N$. 
  
We propose a quasi-binomial mixed effect model to describe the relationship between methylated counts, $S_{ij}$ for $j = 1, 2, \ldots m_i$, and the sample-level covariates $\bs{Z}_i$. Specifically,
\begin{eqnarray}
\label{model1}
\log \dfrac{\pi_{ij}}{1-\pi_{ij}} &=& \beta_0(t_{ij}) + 
\beta_1(t_{ij})Z_{1i} + \beta_2(t_{ij})Z_{2i} + \ldots + \beta_P(t_{ij})Z_{Pi} + u_i,    \\
u_i &\overset{iid}{\sim} & N(0,  \sigma_0^2)  \notag \\
\mathbb{V}\text{ar}(S_{ij} \mid u_i)
&=& 
\phi   X_{ij} \pi_{ij} (1-\pi_{ij})   %	S_{ij} \sim \text{QuasiBinomial} (X_{ij}, \pi_{ij}, \phi) 
\label{conditiona_var}
\end{eqnarray}	
where $\pi_{ij} = \mathbb{E} \left( S_{ij} \mid u_i \right)/X_{ij}$ is the \textit{individual's} methylation proportion (i.e. the conditional mean),  $\beta_0(t_{ij})$ and $\left\{\beta_p(t_{ij})\right\}_{p =1}^P$ are functional parameters for the intercept and covariate effects on $\pi_{ij}$, and $\sigma^2_0$ is the random effect variance.
In this model, we assume the underlying proportion of methylated reads for the $i^{th}$ sample at the $j^{th}$ CpG site, $\pi_{ij}$, depends on covariates $\bs{Z}_i$ and on nearby methylation patterns through a logit link function. In addition, each $\pi_{ij}$ incorporates a subject-specific random intercept (i.e. an additive overdispersion) $u_i$ that is normally distributed and independent across samples. The inclusion of $u_i$ allows for sample heterogeneity in baseline methylation patterns, and at the same time accounts for the correlation among methylation measurements taken on the same sample. Moreover, we assume the variance of $S_{ij}$ for individual samples to be a product of a multiplicative dispersion parameter $\phi$ and a known mean-variance function implied by a binomial distribution ($V(\pi_{ij}) = X_{ij} \pi_{ij} (1-\pi_{ij})$). 
%To allow for the two types of extra-binomial variation, we introduce a subject specific random effects $u_i$ and a dispersion parameter $\phi$. The latter imposes a proportional adjustment on the conditional mean-variance relationship given the values of random effects $u_i$. We proposed the following model

%$\pi_{ij} = P(S_{ijk} =1 \mid u_i)$ is the unknown methylation proportion parameter for individual subjects, 

%increasing read-depth does not reduce the variability introduced by $\phi$

Both the random effects $\bs{u} = (u_1, u_1, \ldots u_N)^T$ and the multiplicative dispersion parameter $\phi$ capture extra-binomial dispersion. However, they address two different aspects of dispersion: $\bs{u}$ models the variation that is due to independent noise across samples, while $\phi$ aims to relax the assumption of the conditional distribution of $S_{ij}$ given $\bs{u}$ such that it is not confined to a binomial distribution.  In fact, our model generalizes the binomial-based model in \cite{zhao2020novel} by introducing both the additive dispersion term $\bs{u}$ and multiplicative dispersion term $\phi$. Specially, imposing $\phi = 1$ in model \eqref{model1} leads to an additive-dispersion-only model and $\sigma_0^2 =0$ corresponds to a multiplicative-dispersion-only model.
When $\sigma_0^2 =0$ and $\phi = 1$, our model reduces to the binomial-based model in \cite{zhao2020novel}.

% special cases, phi = 1, sigma= 0 
% phi =1, sigma>0 --- conditiona distribution if binomial - additive dispersion-only model
% phi >=1, sigma =0 --- between-subject heterogeneity can be ignored -- mutliplicative dispersion-only model

\subsubsection{Marginal interpretations}
A key feature of the mixed effect model in \eqref{model1} is that the regression coefficients $\beta_p(t_{ij})$ need to be interpreted conditional on the value of random effect $u_i$. For example, $\beta_p(t_{ij})$ describes how an \textit{individual's} methylation proportions in a region depend on covariate $Z_p$. If one desires estimates of such covariate effects on the average population, it is more appropriate to determine the marginal model implied by \eqref{model1}.  After applying a cumulative Gaussian approximation to the logistic function and taking an expectation over $u_i$, it can be shown that the marginal mean, $\pi_{ij}^M$, has the form
\begin{equation}
\label{marginal_mean}
\pi_{ij}^M = 
\mathbb{E}(S_{ij})/X_{ij}
\approx  g \left( \sum_{p=0}^P a\; \beta_p(t_{ij})Z_{pi}\right),
\end{equation}
where $g(x) =1/\left(1+\exp(-x)\right)$, $Z_{0i} \equiv 1$, and the constant $a = (1+ c^2\sigma_0^2)^{-1/2} $ with $c= \sqrt{3.41}/\pi$; see detailed derivations in Appendix \ref{deriv_mean}.  The approximation in \eqref{marginal_mean} is quite accurate with errors  $\leq0.001$. Thus, the marginal mean induced by our mixed effect model depends on the covariates $Z_p$ through a logistic link with attenuated regression coefficients $a\beta_p(t_{ij})$. Although the smooth covariate effect parameters $\beta_p(t_{ij})$ have no marginal interpretation, they do have a strong relationship to their marginal counterparts. Hence, the results from hypothesis testing $H_0: \beta_p(t_{ij})=0$ describe the significance of the covariate effect on both the population-averaged and an individual's DNA methylation levels across a region.

Similarly, the marginal variance of $S_{ij}$ does not coincide with its conditional counterpart as shown in \eqref{conditiona_var}. Specifically, our mixed effect model implies a marginal variance of $S_{ij}$ defined as
\begin{eqnarray}
\label{marginal_var}
\mathbb{V}\text{ar}(S_{ij}) &\approx& X_{ij} \pi_{ij}^\star(1-\pi_{ij}^\star)\left\{\phi + \sigma_0^2\left( X_{ij} - \phi \right) \pi_{ij}^\star(1-\pi_{ij}^\star) \right. \notag \\
 &&+ \left. \sigma_0^2/2(1-2\pi_{ij}^\star)^2\left[ 1+ \sigma_0^2\pi_{ij}^\star (1-\pi_{ij}^\star)(X_{ij} -\phi - 1/2)\right] \right\},
%\mathbb{V}\text{ar}(S_{ij}) \approx X_{ij} \pi_{ij}^\star(1-\pi_{ij}^\star)\left\{\phi + \sigma_0^2 X_{ij} \pi_{ij}^\star(1-\pi_{ij}^\star) \right\},
\end{eqnarray}
where $\pi_{ij}^\star = g^{-1} \left(  \sum_{p=0}^P \beta_p(t_{ij})Z_{pi} \right)$; see detailed derivations in Appendix \ref{deriv_var}. Note that $\pi_{ij}^\star$ is the mean methylation proportion when setting random effects $u_i$ to zero and is related to the marginal mean $\pi_{ij}^M$ via 
$\pi_{ij}^\star = g\left( g^{-1}\left( \pi_{ij}^M \right)/a \right)$. Equation \eqref{marginal_var} illustrates that, under the dSOMNiBUS model, the marginal variance of methylated counts at a CpG site is approximately the variance of the binomial model multiplied by a dispersion factor $\phi^\star=  \phi + \sigma_0^2\left( X_{ij} - \phi \right) \pi_{ij}^\star(1-\pi_{ij}^\star)+  \sigma_0^2/2(1-2\pi_{ij}^\star)^2\left[ 1+ \sigma_0^2\pi_{ij}^\star (1-\pi_{ij}^\star)(X_{ij} -\phi - 1/2)\right]$, which depends on the combined effect of $\phi$, the multiplicative dispersion for the conditional variance given the RE, and $\sigma_0^2$, the variance of the subject-level RE. Notably, the marginal dispersion factor $\phi^\star$ also depends on genomic position $t_{ij}$ via the dependence of $\pi_{ij}^\star$ on $t_{ij}$. Consequently, our dSOMNiBUS model in \eqref{model1} naturally allows dispersion levels to vary across loci, whereas a multiplicative-dispersion-only model (i.e. $\sigma_0^2 = 0$) can only accommodate constant dispersion in a region, as illustrated in Figure \ref{fig:fig2-dispersionpatternbeta1actual}. It is also clear from Equation \eqref{marginal_var} that an additive-dispersion-only model (i.e., $\phi =1$) only allows for overdispersion, and the combination of additive and multiplicative dispersion naturally accounts for both over- and underdispersion.
% consider to add a small section summarizing the inference/algorithm or put all the details in the method section?
 \subsubsection{Dealing with possible measurement errors in methylated counts}
In the presence of experimental errors, the true methylation data, $S_{ij}$ are unknown and one only observes $Y_{ij}$.
We assume the following error mechanism
\begin{eqnarray}
\label{error}
P(Y_{ijk} = 1 \mid S_{ijk} = 0 )&=&  p_0  \notag \\
P(Y_{ijk} = 1 \mid S_{ijk} = 1 )&=&  p_1. 
\end{eqnarray}
Here, these two parameters capture errors; $p_0$ is the rate of false methylation calls, and $1-p_1$ is the rate of false non-methylation calls. 
These rates are assumed to be constant across all reads and positions. The error parameters $p_0$ and $p_1$ can be estimated by looking at raw sequencing data at CpG sites known in advance to be methylated or unmethylated \citep{wreczycka2017strategies}. We assume hereafter that $p_0$ and $p_1$ are known.
The methodology details on how to make inference about covariate effects $\beta_p(t_{ij})$ and estimate dispersion parameters $\phi$ and $\sigma_0^2$, in the presence of data errors,  are described in Section \ref{estimation_with_error}.
% ASENtence is missing here showing how dSOMNiBUS incorporates these parameters. Perhaps pointing to an appendix. --- already added
\subsection{Illustration of performance of dSOMNiBUS in the ACPA dataset} \label{data}
We first apply our approach to targeted bisulfite sequencing data from a rheumatoid arthritis study \citep{shao2019rheumatoid}.   Participants were sampled from the CARTaGENE cohort (\url{https://www.cartagene.qc.ca/}), a population-based cohort including 43,000 general population subjects aged 40 to 69 years in Quebec, Canada. The study aims to investigate association between DNA methylation and the levels of anti-citrullinated protein antibodies (ACPA), a marker of rheumatoid arthritis (RA) risk that often presents prior to any clinical manifestations \citep{forslind2004prediction}. 

Firstly, the serum ACPA levels were measured for a randomly sampled 3600 individuals from the CARTaGENE cohort, based upon which individuals were classified as either ACPA positive or ACPA negative. Then, the whole blood samples of the  ACPA positive individuals, and a selected subset of  age-sex-and-smoking-status-matched ACPA negative individuals were sent for Targeted Custom Capture Bisulfite Sequencing.  Specifically, the sequencing used an immune targeted panel that covers the majority of genomic regions with relevance to RA and blood cells. 
%NEEDs improvement is there a better phrase in Shao et al. about the panel design
Cell type proportions in the blood samples were also measured at the time of the sampling \citep{shao2019rheumatoid}. 

Using this sampling approach, two batches of data, referred to as data 1 and data 2, were collected in 2017 and 2019, respectively. Notably, the classification criteria for ACPA status are slightly different between data 1 and 2.  When sampling data 1, subjects with serum ACPA levels greater than 20 optical density (OD) units were called as ACPA postive and samples with ACPA levels less than 20 OD were defined as ACPA negative. After data cleaning, data 1 consisted of 69 ACPA positive subjects and 68 ACPA negative subjects. In contrast, the sampling of data 2 was based on more extreme cutoffs for ACPA levels, and resulted in 60 ACPA positive subjects (ACPA levels $\geq$ 60 OD) and 60 ACPA negative subjects (ACPA levels $< 20$ OD). This change in decision is reflected in the different distributions of serum ACPA levels between data 1 and 2, as shown in Supplementary Figure S1.  Average sequence read depths in targeted regions were 5 and 35 in data 1 and 2, respectively (Supplementary Figure S2), due to improvement in the sequencing protocols implemented between the two experiments. % Read-depth comparison

%after adjustment for age, sex, smoking status and cell type composition. 
%Overall, the methylation data cover $\sim 4,860,000$ CpG sites, allocated to $\sim400,000$ regions. 
In this article, we restricted our attention to regions with at least 50 CpG sites. In addition, we excluded regions with more than 95\% CpGs having median read depth 0  or having median methylation proportion as 0. Overall, we analyzed 10,759 regions in dataset 1 and 12,983 regions in dataset 2. 
We excluded the samples who reported a diagnosis of RA before the CARTaGENE study started. Subjects with missing information on cell type proportions were also removed from our analysis. Supplementary Table S1 presents the sample characteristics in data 1 and 2. 

We apply our approach to both data 1 and 2, with the aim to identify the differentially methylated regions that show association with ACPA, after adjustment for age, sex, smoking status and cell type composition. Specifically, we assumed no data errors in the datasets ($p_0 = 1- p_1 = 0$). We used natural cubic splines to expand the smooth terms in the model, and its rank $L_p$ was approximately as the number of CpGs in a region divided by 10 for $\beta_0(t)$, and divided by 20 for $ \beta_p(t), p\geq 1$. 
%evaluate the impact of ACPA status on methylation patterns in all targeted regions, with adjustment for age, sex, smoking status and cell type composition.
%We aim to identify the differentially methylated regions that show association with ACPA, after adjustment for age, sex, smoking status and cell type composition, in the individuals without clinical manifestations of RA.

\subsubsection{Both additive and multiplicative dispersion is present in the data}
  
\begin{figure}[h!]
\centering
(A)\\
\includegraphics[width=0.8\linewidth]{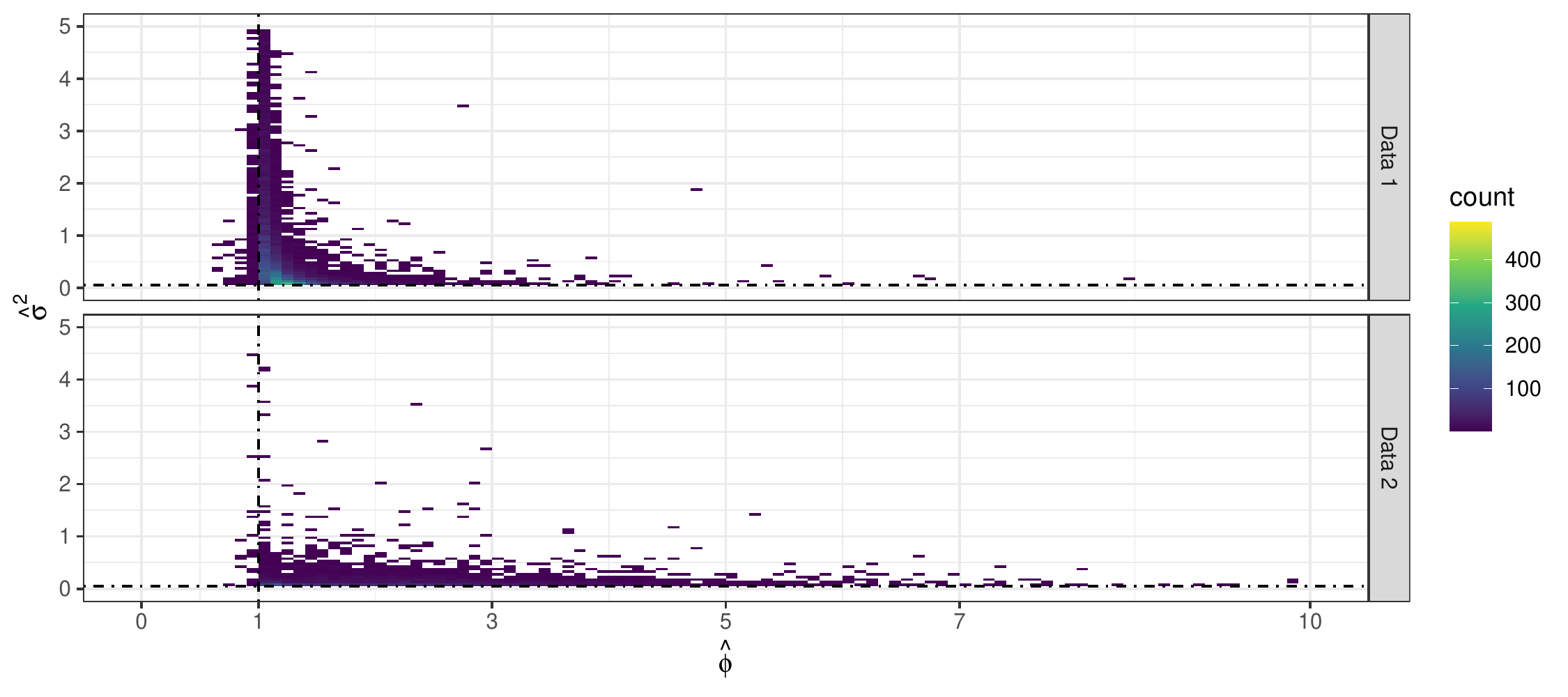}\\
\centering
(B) \quad \qquad  \qquad \qquad  \qquad \qquad  \qquad (C)\\
\includegraphics[width=0.4\linewidth]{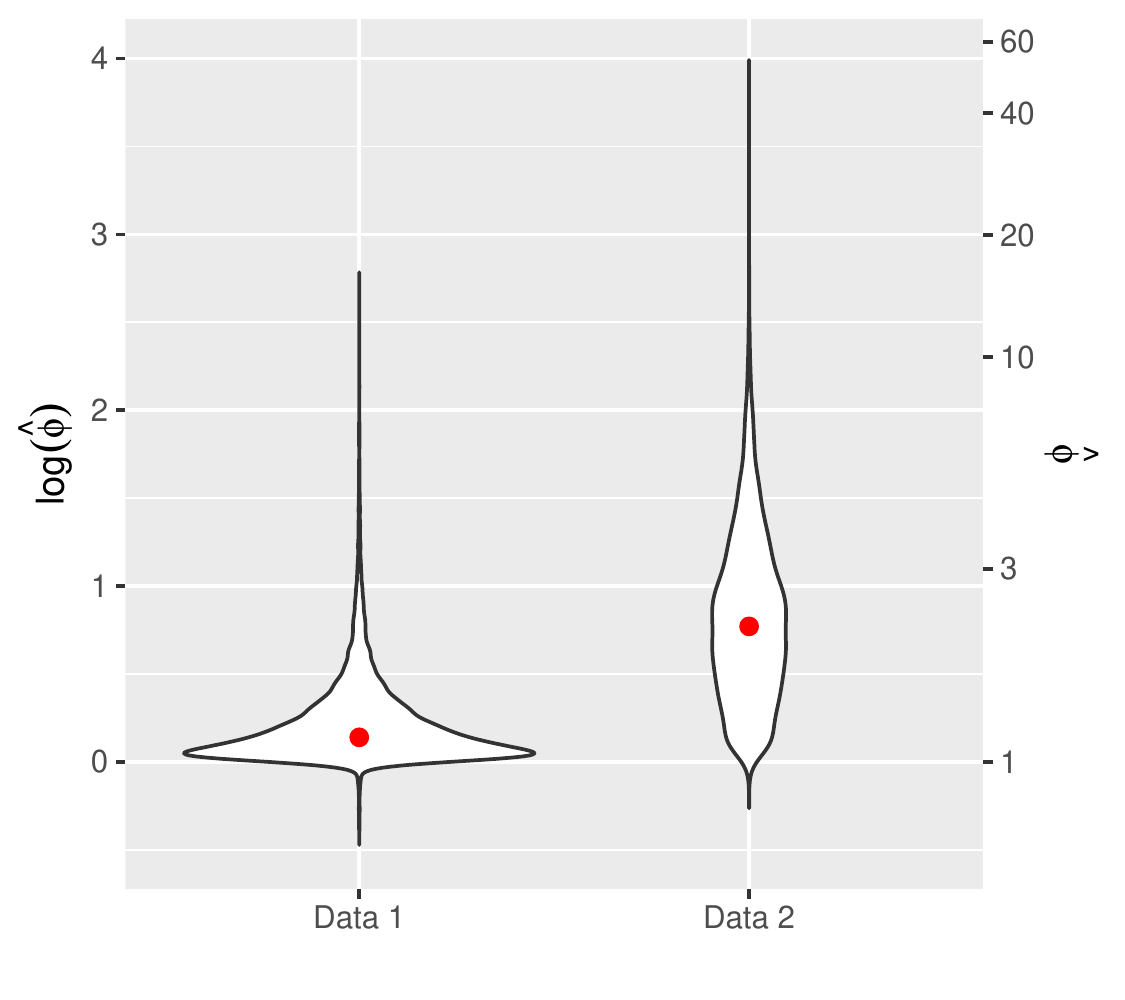}
\includegraphics[width=0.4\linewidth]{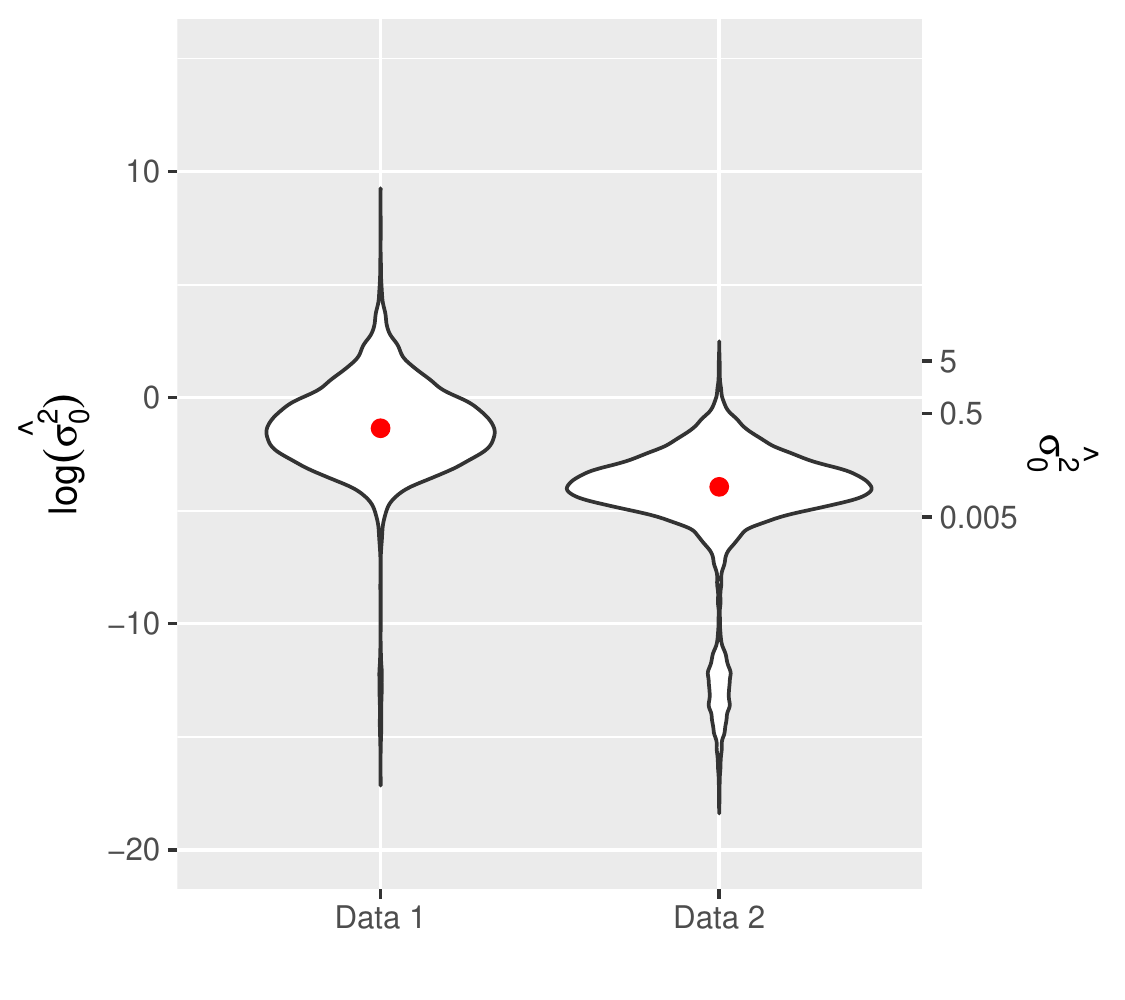}
\caption{Distribution of the estimated multiplicative dispersion parameter $\phi$ and additive dispersion parameter $\sigma_0^2$, for all test regions in dataset 1 and 2. Panel (A) shows the 2-dimensional histogram for $\widehat{\phi}$ and $\widehat{\sigma}_0^2$, where the color intensity represents the number of regions with a particular combination of values of  $\widehat{\phi}$ and $\widehat{\sigma}_0^2$. Panels (B) and (C) show the rotated kernel density plots (i.e. violin plots) for $\widehat{\phi}$ and $\widehat{\sigma}_0^2$ (in a natural logarithmic scale), separately.}
\label{fig:realdata-phi-sigma0-combined}
\end{figure}

Figure \ref{fig:realdata-phi-sigma0-combined} presents the distribution of estimated multiplicative dispersion $\phi$ and additive dispersion $\sigma_0^2$ for all test regions in dataset 1 and 2. Overall, widespread overdispersion is observed; 98.5\%  regions show multiplicative dispersion $\phi$ greater than 1 and 51.2\% regions show additive dispersion $\sigma_0^2$ greater than 0.05. The Pearson correlation coefficient between the estimated $\phi$ and $\sigma_0^2$ is $-0.015$. There exist 49.8\% regions with both multiplicative dispersion $\phi > 1$ and additive dispersion $\sigma_0^2 >0.05$.
% Folder "SOMNiBUS_RE_Simu/Summary_real_data_results"
% More person-to-person variation is observed in data 1

\subsubsection{Ignoring either type of dispersion leads to inflated type I errors}
\begin{figure}[h!]
\centering
\includegraphics[width=0.9\linewidth]{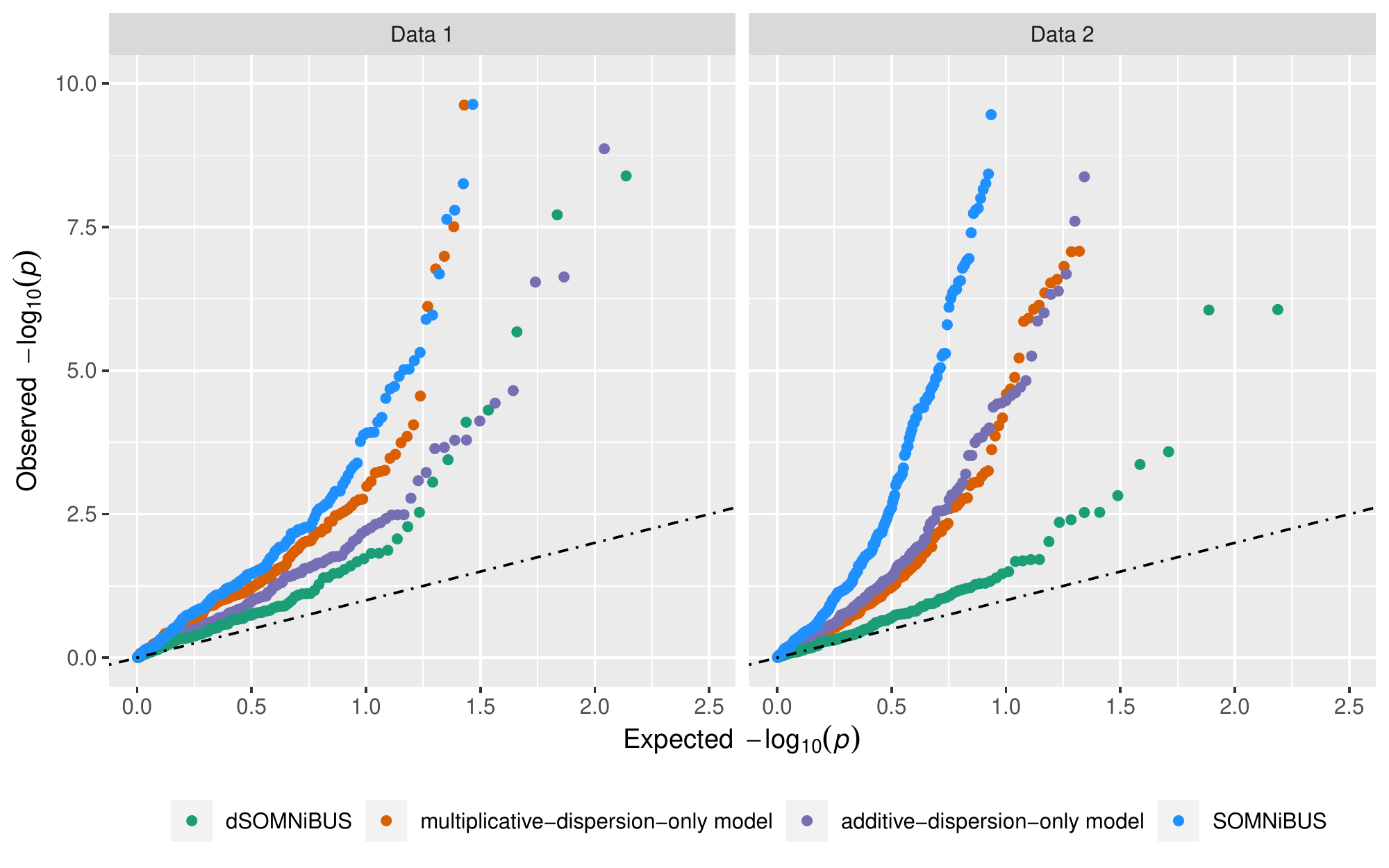}
\caption{QQ plot for regional p-values, obtained from models addressing different types of dispersion.}
\label{fig:excessheterogeneity}
\end{figure}

Figure \ref{fig:excessheterogeneity} shows quantile-quantile (QQ) plots for the regional p-values for the effect of ACPA on the 292 regions of Chromosome 18 in the two datasets. Detailed inference steps are given in Section \ref{method_detail}. The results are compared among four different approaches: (1) dSOMNiBUS which models both the multiplicative and additive dispersion, (2) the multiplicative-dispersion-only model, (3) the additive-dispersion-only model, and (4) the standard SOMNiBUS which ignores any extra-binomial variation. Figure \ref{fig:excessheterogeneity} reveals that, when ignoring either type of dispersion, the distribution of regional p-values is biased away from what would be expected under the null. The inclusion of both multiplicative and additive dispersion is important for correct type I error control.

%Figure \ref{fig:excessheterogeneity} shows the results for 292/298 regions on Chromosome 18 for data 1 and 2, obtained from four approaches --- 1) `No dispersion; No RE`, model \eqref{model1} with $\sigma_0^2 =0$ and $\phi =1$, 2) `RE-only' model \eqref{model1} with $\sigma_0^2 > 0$ and $\phi =1$, 3) `Dispersion only` model \eqref{model1} with $\sigma_0^2 = 0$ and $\phi >1$, and 4) `Dispersion $+$ RE`, model \eqref{model1}.

\subsubsection{Our inference procedure provides well-calibrated p-values}
\begin{figure}[h!]
\centering
\includegraphics[width=0.9\linewidth]{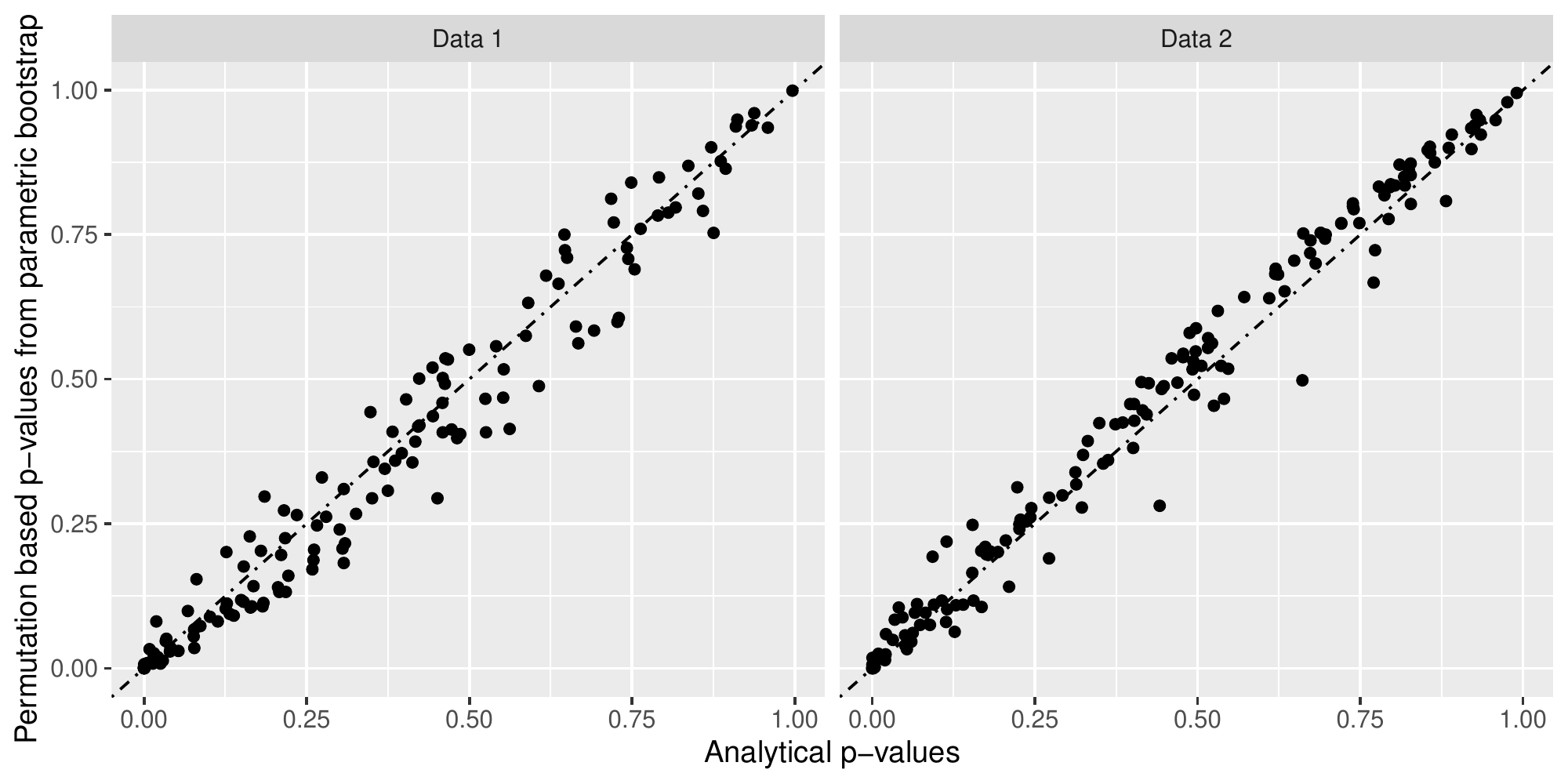}
\caption{Comparison between the observed regional p values from our approach and the permulation-based p values from parametric bootstrap.}
\label{fig:permutation}
\end{figure}

%We also use the RA datasets to test the validity of our proposed inference.
To test DMRs, we propose a region-based statistic with a F limiting distribution; see details in Section \ref{test}. To test the validity of our inference, we compare our regional p-values to bootstrap-based p-values, whose null distribution is constructed by parametric bootstraps   \citep{{davison1997bootstrap}} and does not rely on any distributional assumptions. 
Figure \ref{fig:permutation} shows the distributions of bootstrap-based and our analytical p-values for the targeted regions on chromosome 18, demonstrating that our inference method generates p-values in line with the bootstrap-based results. Thus, dSOMNiBUS provides accurate tests for DMRs without requiring extensive computational time.

%accurate inference results.
% provides p-values in line with null expectation for this region.

%To obtain statistical inference without making strong distributional assumptions, 

%We also performed permutation tests to demonstrate that the proposed inference of smooth covariate effects in Section XXX is valid, we compare our analytical p-values to the permutation-based p-values. Specifically, 

%Our method uses a region-based statistic with a F limiting distribution. We also compared our inference to permutation-based results, where the null distribution was constructed by parametric bootstrap, other than F distribution.

\subsection{Simulation study}
We conducted simulations to assess the proposed inference of smooth covariate effects, and to compare the performance of our method with five existing methods: \texttt{BiSeq} \citep{hebestreit2013detection}, \texttt{BSmooth} \citep{hansen2012bsmooth}, \texttt{SMSC} \citep{lakhal2017smoothed}, \texttt{dmrseq} \citep{korthauer2017detection} and \texttt{GlobalTest} \citep{goeman2006testing}, in terms of type I error and power. Detailed descriptions of these five methods are given in 
Supplementary Section 3.2. We also made special modifications for the implementations of \texttt{BSmooth}, \texttt{SMSC} and \texttt{dmrseq}, which are primarily designed for WGBS data, to make them as appropriate as possible for targeted regions. see details in Supplementary Section 3.1.

\subsubsection{Simulation design}
We adopt similar simulation parameters as described in \cite{zhao2020novel}, and simulated methylation regions with 123 CpG sites under various settings. We first generated the read depth $X_{ij}$ by adding Bernoulli random variables (with proportion 0.5) to a pre-specified regional read-depth pattern (Supplementary Figure S3). 
% THIS is unclear, adding bERNOULLI to what
In this way, the spatial correlation of read depth observed in real data was well preserved in the simulated data. The rest of simulation parameters were defined  in Table \ref{tab:my-table}.

%We then considered the following two scenarios to specify covariates ${Z}_p$ and their effect curves $\beta_p(t)$

%All simulation parameters are summarized in Table \ref{tab:my-table}

\begin{table}[h!]
	\caption{Simulation settings for the functional parameters $\beta_p(t)$, sample size $N$, error parameters $p_0$ and $p_1$, multiplicative parameter $\phi$ and RE variances $\sigma_0^2$.}
	\label{tab:my-table}
	\scalebox{0.8}{
		\begin{tabular}{ll}
			\hline
			Simulation  & Possible values \\
			parameters                                                                                     \\ \hline
			$\beta_p(t)$                        &   Scenario 1: three covariates: $Z_1\sim Bernoulli(0.51)$, $Z_2 \sim Bernoulli(0.58)$ and $Z_3\sim Bernoulli (0.5)$                                                                                                  \\
			&  \qquad  with effects $\beta_1(t), \beta_2(t)$ and $\beta_3(t)$ and intercept $\beta_0(t)$, shown in the red curves in Figure \ref{fig:finalshadow-plotphi3re3quasire}. \\
			& \qquad Here, $Z_3$ is the null covariate with effect $\beta_3(t) \equiv 0$. \\
			& Scenario 2: one covariate: $Z\sim Bernoulli(0.5)$                \\
			& \qquad with 15 different settings of $(\beta_0(t), \beta_1(t))$, which yield methylation proportion parameters \\
			& \qquad as depicted in Figure \ref{fig:power-setting-finner}.        \\
			$N$                              & 100                                                                               \\
			$(p_0, p_1)$                          & $(0.003, 0.9)^{\dagger}$ or $(0, 1)$                                                                                  \\
			$\phi$ & $(1, 3)$ \\
			$\sigma_0^2$ & $(0, 1, 3, 9)$, and the corresponding subject-specific RE $u_i \overset{i.i.d}{\sim} N(0, \sigma_0^2)$ for $i = 1, 2, \ldots N$\\
			%		& Sensitivity analysis in Section \cite{errorp} $p_0 = $
			%	Methods                        & (\texttt{SOMNiBUS}, \texttt{BiSeq}, \texttt{dmrseq},	\texttt{BSmooth}, \texttt{SMSC}  \\     
			\hline
			\multicolumn{2}{l}{$\dagger$ the value $0.003$ was reported by \cite{prochenka2015cautionary} as insufficient Bisulfite conversion rate and $0.1$ was}\\
			\multicolumn{2}{l}{\; estimated as the average excessive conversion rate from a (single-cell-type) bisulfite dataset in }\\
			\multicolumn{2}{l}{\; \cite{hudson2017novel} using the method SMSC \citep{lakhal2017smoothed}.}
		\end{tabular}
	}
\end{table}

%\[
%\omega \pi_0 + (1-\omega)\pi_1,\; \text{with } \omega = (0, 1/14, 2/14, \ldots 13/14, 1)
%\]
\begin{figure}[h!]
	\centering
	\includegraphics[width=0.7\linewidth]{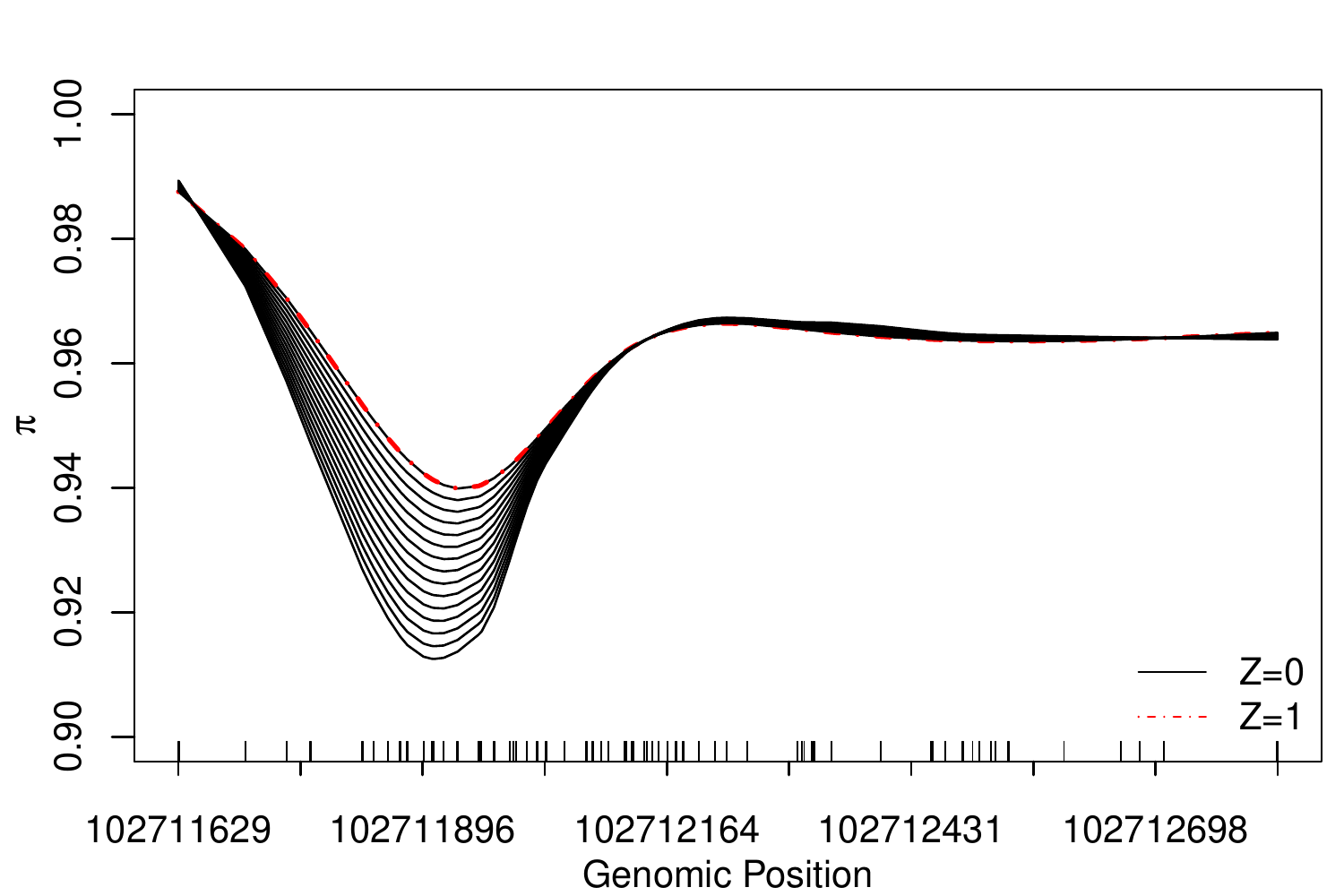}
	\caption{The 15 simulation settings of methylation parameters $\pi_0(t)$ and $\pi_1(t)$ in Scenario 2. Here, $\pi_0(t)$ and $\pi_1 (t)$ denote the methylation parameters for samples with $Z = 0$ and $Z = 1$ at position $t$, respectively. Under this scenario, $\pi_1(t)$ (red dotted-dashed curve) is fixed across settings, whereas $\pi_0(t)$s (black solid lines) vary across settings corresponding to different degrees of closeness between methylation patterns in the two groups.
	}
	\label{fig:power-setting-finner}
\end{figure}

\paragraph{Simulate dispersed-binomial counts.}
Given the values of $\{Z_1, \ldots Z_P\}$, $\{\beta_p(t), p = 0, 1, \ldots P\}$ and $\{u_i, i = 1, 2, \ldots N\}$ under each setting, the individual's methylation proportion, $\pi_{ij}$, can be readily calculated from the mean model in \eqref{model1}. We then generated the true methylation counts $S_{ij}$ from a beta-binomial distribution with proportion parameter $\mu = \pi_{ij}$, correlation parameter $\rho = \dfrac{\phi -1}{X_{ij} -1}$, and size parameter $n = X_{ij}$. Specifically, $S_{ij}$ were drawn from the following probability mass function
\[
P(S_{ij} = k\mid \mu, \rho, n) = \left( \begin{matrix*}
n \\ k
\end{matrix*}\right) \dfrac{B(k+\alpha, n-k+\beta)}{B(\alpha, \beta)}
\]
where $\alpha = \mu(1-\rho)/\rho, \beta = (1-\mu)(1-\rho)(1-\mu)/\rho$, and $B(\cdot,\cdot)$ is the beta function. The variance of $S_{ij}$ can be thus derived as
\[
\mathbb{V}\text{ar}(S_{ij}) = \left[1+(n-1)\rho  \right] \left[  n \mu (1-\mu) \right] = \phi X_{ij} \pi_{ij} (1-\pi_{ij}),
\]
which coincides with our assumed mean-variance relationship in \eqref{conditiona_var}. We then generated the observed methylated counts $Y_{ij}$ according to the error model in \eqref{error}, which implies
$$
Y_{ij} \mid S_{ij} \sim \text{Binomial} (S_{ij}, p_1) + \text{Binomial} (X_{ij} - S_{ij}, p_0).
$$
%The simulated counts will then be fed into the statistical model described in Method

%We considered two settings for error parameters $p_0$ and $p_1$: (1) $p_0 = 0.003$ and $1-p_1 = 0.1$, and (2) $p_0 = 1-p_1 = 0$.
%The values for $p_0$ and $p_1$ in the first setting were selected because they have been reported as incomplete and over conversion rates in a bisulfite sequencing experiment conducted by \cite{prochenka2015cautionary}.
%The simulated counts will then be fed into the statistical model described in Method

Under each scenario and setting, we generated data sets with sample sizes $N = 100$, each 1000 times.
We then applied \texttt{dSOMNiBUS} along with methods \texttt{BiSeq}, \texttt{dmrseq}, \texttt{BSmooth},  \texttt{SMSC} and \texttt{GlobalTest} to the simulated data sets. 
%Unless otherwise stated, default settings were used for the five alternative methods. 
For our approach \texttt{dSOMNiBUS}, we used cubic splines with dimension $L_p = 5$ to parameterize the smooth terms of interest. 
We also assumed that the correct values of error parameters $p_0$ and $p_1$ were known.% although we conducted sensitivity analyses to this assumption (see Discussion and Supplementary Section XXXX).

\subsubsection{dSOMNiBUS provides accurate inference for smooth covariate effects}
\begin{figure}[h!]
	\centering
	\includegraphics[width=1\linewidth]{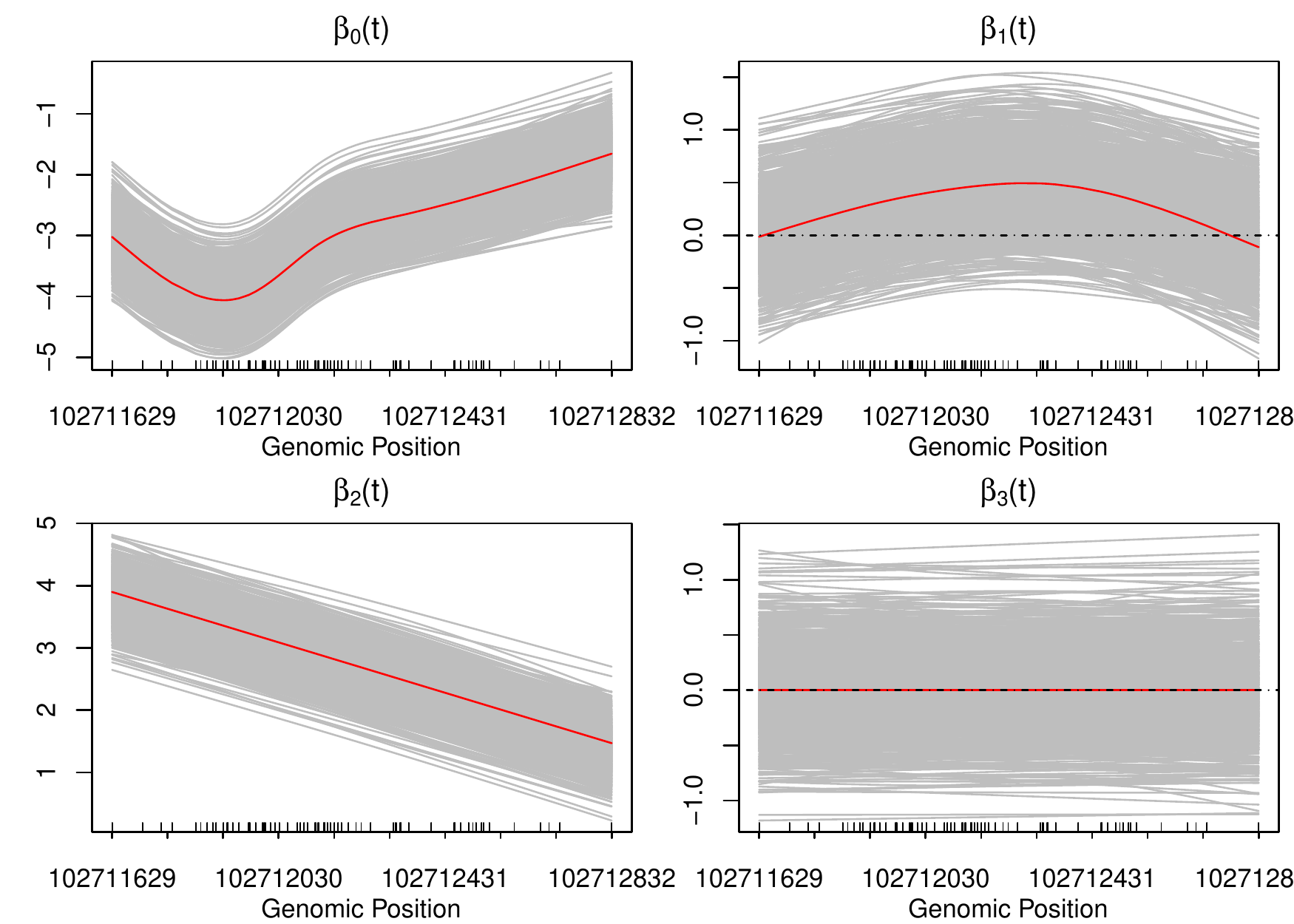}
	\caption{Estimates of smooth covariate effects (gray) over the 1000 simulations in Scenario 1, using dSOMNiBUS. The red curves are the true functional parameters used to generate the data. Data were generated with error using $\phi =3$ and $\sigma_0^2 =3$.}
	\label{fig:finalshadow-plotphi3re3quasire}
\end{figure}
Figure \ref{fig:finalshadow-plotphi3re3quasire} presents the estimates of the functional parameters $\beta_0 (t), \beta_1 (t), \beta_2 (t)$ and $\beta_3 (t)$ over 1000 simulations, obtained from dSOMNiBUS; here, data were generated under Scenario 1, with multiplicative dispersion parameter $\phi =3$, RE variance $\sigma_0^2 =3$, and  error parameters $p_0 = 0.003$ and $ 1-p_1 = 0.1$. Figure \ref{fig:finalshadow-plotphi3re3quasire} demonstrates that the proposed method provides unbiased curve estimates for smooth covariate effects when the regional methylation counts exhibit extra-parametric variation and are measured with errors.
%can provides unbiased curve estimates for all the four functional parameters in the model, 

\begin{figure}[h!]
	\centering
	\includegraphics[width=1\linewidth]{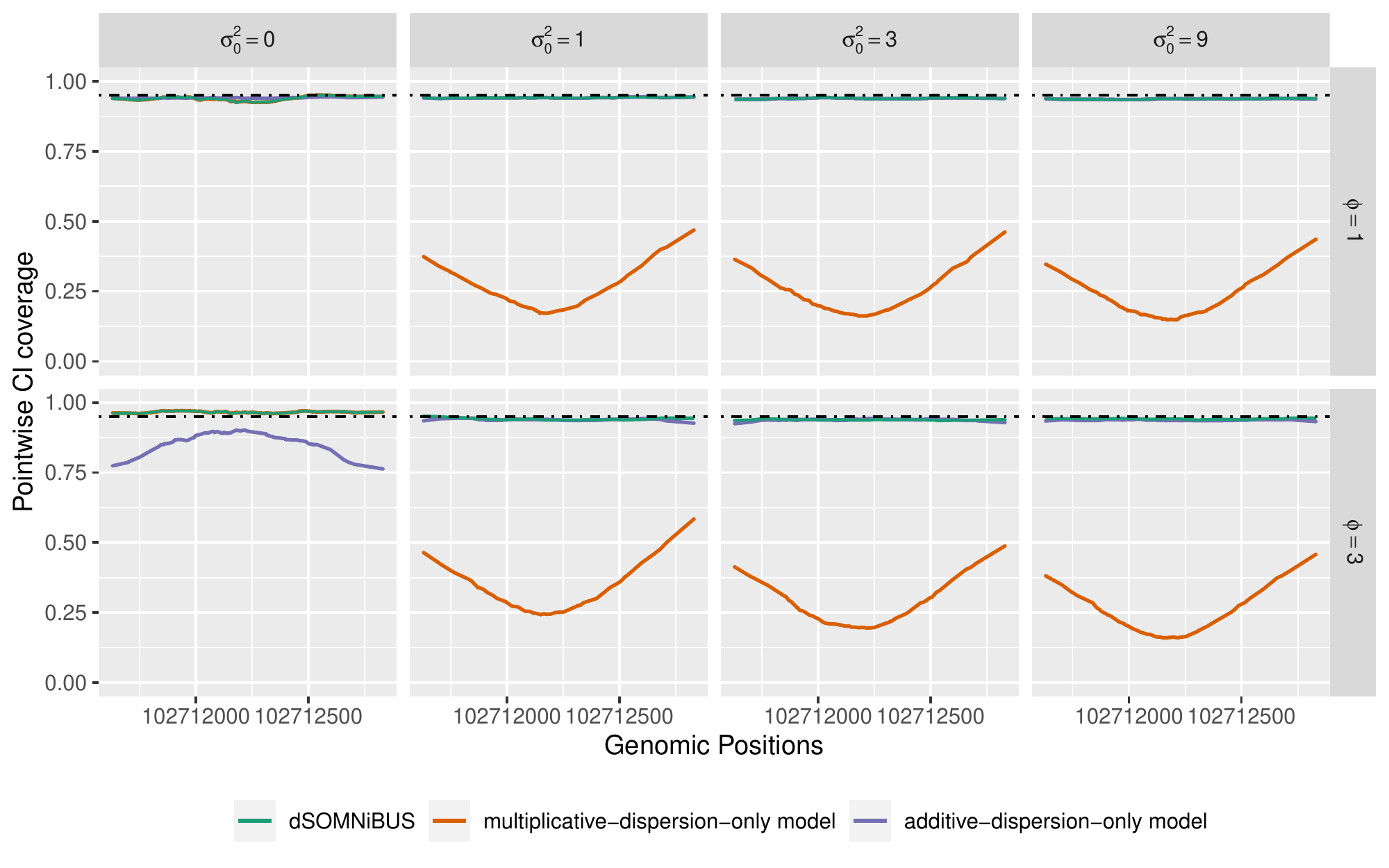}
	\caption{Empirical coverage probability of the analytical 95\% CIs for $\beta_3(t)$ over 1000 simulations, under different vales of $\phi$ and $\sigma_0^2$. The empirical coverage probabilities are defined as the percentage of simulations where the analytical CIs cover the true value of $\beta_3(t)$. Data were generated with error, under simulation Scenario 1. The results from dSOMNiBUS (green) and the additive-dispersion-only model (purple) are indistinguishable in all settings but $\sigma_0^2=0$ and $\phi = 3$ and dSOMNiBUS (green) and the multiplicative-dispersion-only model (orange) are indistinguishable when $\sigma_0^2=0$.}
	\label{fig:2-three-models}
\end{figure}
\begin{figure}[h!]
	\centering
	\includegraphics[width=1\linewidth]{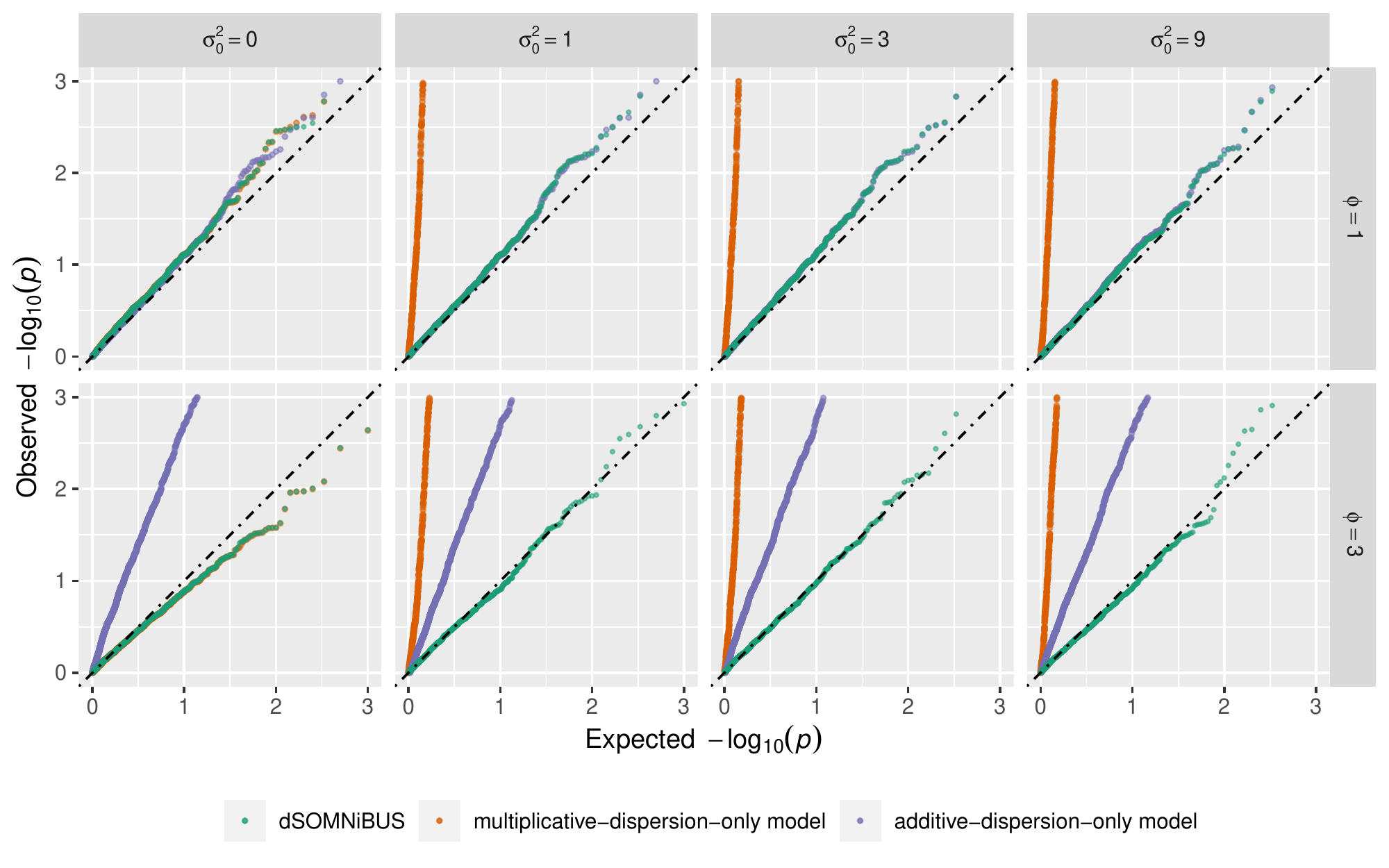}
	\caption{QQ plot for regional p-values for the test $H_0: \beta_3(t) = 0$, obtained from dSOMNiBUS, the multiplicative-dispersion-only model and the additive-dispersion-only model. Data were simulated with error, under simulation Scenario 1. When $\phi = 1$, the results from dSOMNiBUS (green) and the additive-dispersion-only model (purple) are indistinguishable. When $\sigma_0^2 = 0$, the lines for the multiplicative-dispersion-only model (orange) and dSOMNiBUS (green) are indistinguishable.
    }
	\label{fig:1-impact-of-dispersionfixedscale}
\end{figure}

Figure \ref{fig:2-three-models} and \ref{fig:1-impact-of-dispersionfixedscale} demonstrate the performance of the proposed pointwise confidence interval (CI) estimates (Section \ref{CI} ) and regional test (Section \ref{test}), respectively. The results from dSOMNiBUS are compared to the multiplicative-dispersion-only model and the additive-dispersion-only model. Figure \ref{fig:2-three-models} displays the empirical coverage probabilities of the analytical 95\% CIs for $\beta_3(t)$, under different settings of $\phi$ and $\sigma_0^2$. Figure \ref{fig:1-impact-of-dispersionfixedscale} shows the QQ plots for the regional p-values when the null hypothesis $H_0: \beta_3(t) = 0$ is correct. The results show that ignoring the presence of additive dispersion (i.e. the multiplicative-dispersion-only model) leads to substantial estimation bias, poor CI coverage probabilities and highly inflated type I errors. Although the additive-dispersion-only model provides relatively accurate pointwise CIs, the distributions of its regional p-values are biased away from what would be expected under the null, when multiplicative dispersion $\phi>1$. Overall, dSOMNiBUS provides pointwise CIs attaining their nominal levels, and region-based statistics whose distribution under the null is well calibrated, regardless of the types and degrees of dispersion that data exhibit.
Similar results were observed when data were generated without error (Supplementary Figures S5 and S6).

%In summary, Figure \ref{fig:coverage}  shows that the coverages of our 95\% confidence intervals attain their nominal values in most of the simulation settings. This suggests that the proposed CI estimation approach quantifies the underlying uncertainty in the smoothed-EM estimates with reasonable accuracy, although it ignores the uncertainty from estimating the smoothing parameters.
\subsubsection{dSOMNiBUS exhibits greater power to detect DMRs while correctly controlling type I error rates}
\begin{figure}[h!]
	\centering
	\includegraphics[width=1\linewidth]{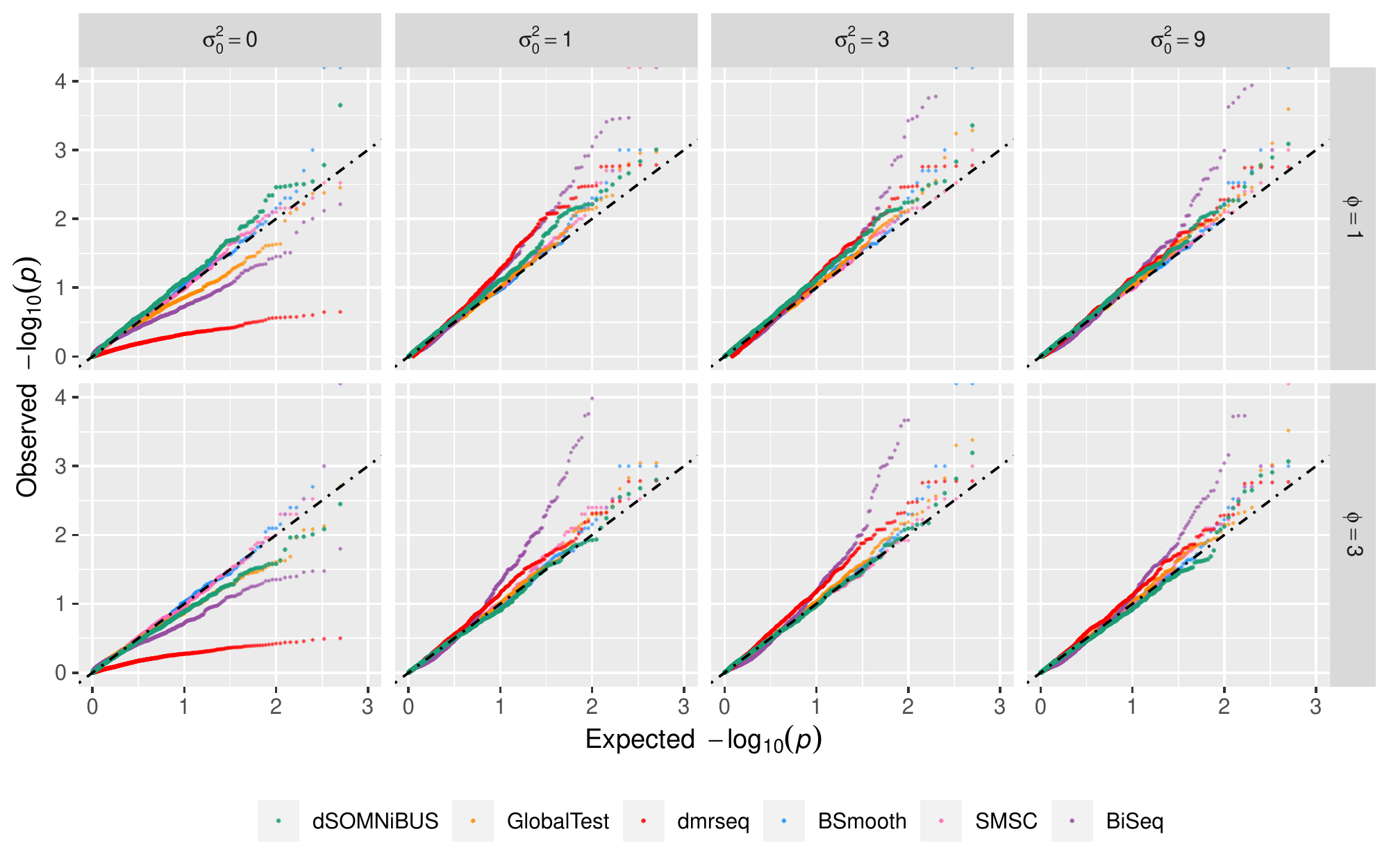}
	\caption{QQ plot for regional p-values for the test $H_0: \beta_3(t) = 0$, obtained from dSOMNiBUS, GlobalTest, dmrseq, BSmooth, SMSC, and BiSeq. Data were simulated with error, under simulation Scenario 1.}
	\label{fig:3-qqplot-6-methods}
\end{figure}

\begin{figure}[h!]
	\centering
	\includegraphics[width=1.1\linewidth]{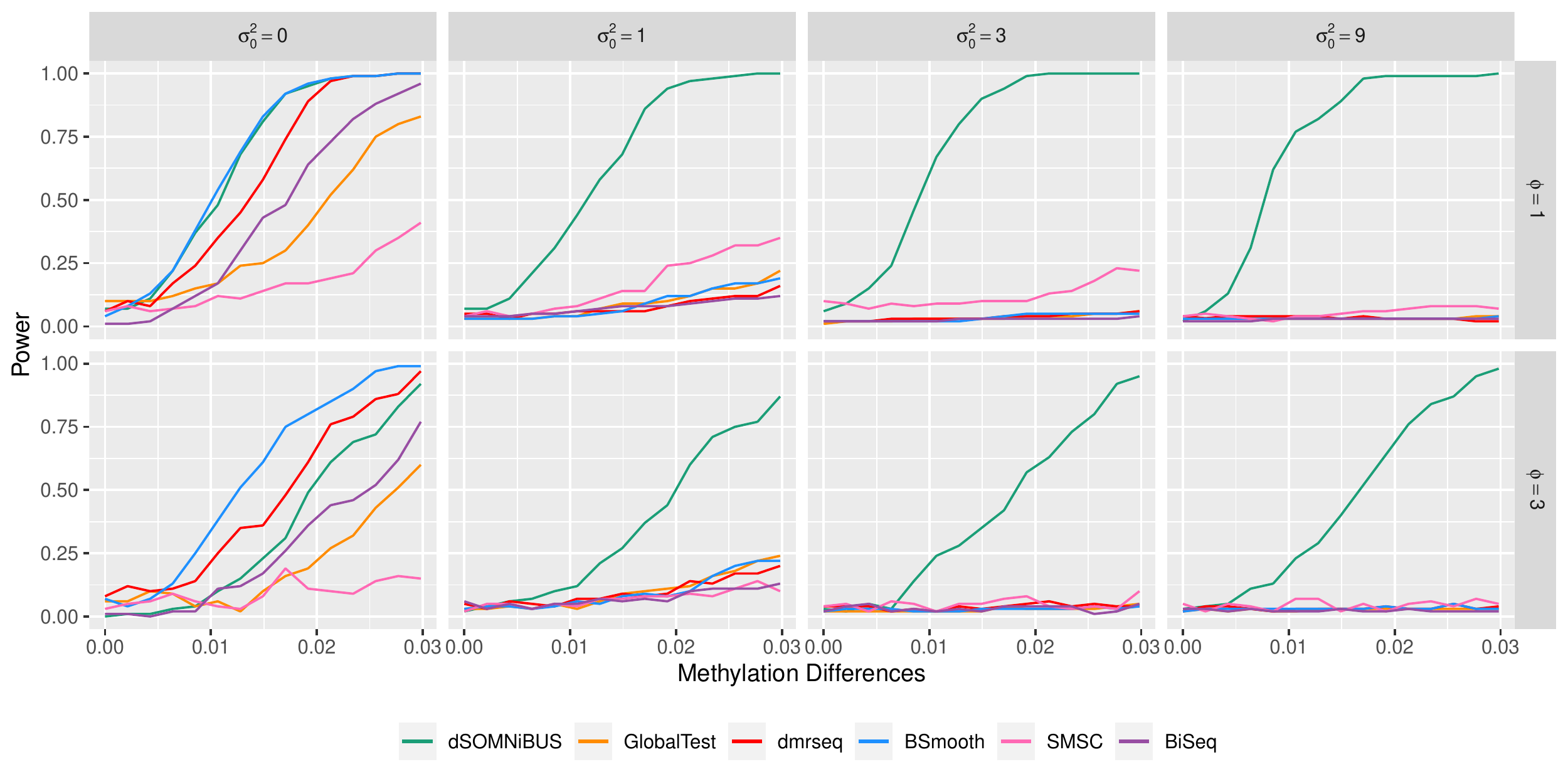}
	\caption{Powers to detect DMRs using the six methods for the 15 simulation settings in Scenario 2 under different levels of maximum methylation differences between $\pi_0(t)$ and $\pi_1(t)$ in the region, calculated over 100 simulations.}
	\label{fig:power}
\end{figure}
Figures \ref{fig:3-qqplot-6-methods} and \ref{fig:power} further demonstrate the performance of the proposed regional test, when compared with the existing methods \texttt{GlobalTest}, \texttt{dmrseq}, \texttt{BSmooth}, \texttt{SMSC}, and \texttt{BiSeq}. 
Here, data were simulated with error parameters $p_0 = 0.003$ and $1-p_1 = 0.1$.
Figure \ref{fig:3-qqplot-6-methods} shows the distributions of p-values for the regional effect of the null covariate $Z_3$. 
Because we estimated the empirical regional p-values for \texttt{BSmooth} and \texttt{SMSC} by permutations, both methods are able to control type I errors, under all settings of $\phi$ and $\sigma_0^2$.
Both \texttt{BiSeq} and \texttt{dmrseq} show deflated type I error rate when $\sigma_0^2 =0$ and inflated type I error rate when $\sigma_0^2 >0$.
The distributions of p-values from \texttt{GlobalTest} are well calibrated when the within subject correlation $\sigma_0^2 >0$, but are slightly biased away from the uniform distribution when $\sigma_0^2 = 0$.
When $\sigma_0^2 = 0$ and $\phi = 3$, dSOMNiBUS provides slightly conservative type I errors; this bias vanishes when the data were generated without error (Supplementary Figures S7).  
Figure \ref{fig:power} shows the powers of the six methods for detecting DMRs under the 15 settings of methylation patterns displayed in Figure \ref{fig:power-setting-finner}. Here, methylation difference is defined as the maximum difference between $\pi_1(t)$ and $\pi_0(t)$ in the region. When data exhibit neither additive nor multiplicative dispersion, dSOMNiBUS and \texttt{BSmooth} provide the highest power, followed by \texttt{dmrseq}, \texttt{BiSeq}, \texttt{GlobalTest}, and \texttt{SMSC}. When $\sigma_0^2 = 0$ and $\phi = 3$,                                                                                                                                                                                                                                                                                                                                                                                                                                                                                                                                                                                                                                                                                                                                                                                                                                                                                                                                                          \texttt{BSmooth} and \texttt{dmrseq} are more powerful than other methods. When there are correlations among methylation measurements on the same subject, i.e. $\sigma_0^2 >0$, dSOMNiBUS clearly outperforms the five alternative methods; this superiority remains when the data were generated without error (Supplementary Figures S8). In summary, dSOMNiBUS exhibits greater power to detect DMRs, while correctly controlling type I error rates, especially when the regional methylation counts exhibit (additive) extra-binomial variation.
%In summary, dSOMNiBUS addresses both multiplicative and additive sources of dispersion in methylation data and makes 

%making reliable inference at the region level.

%In summary, dSOMNiBUS exhibits greater power to detect DMRs than the five alternative , while correctly controlling type I error rates
\section{Discussion}
We have proposed and evaluated a novel method, called dSOMNiBUS, for estimating smooth covariate effects for BS-seq data. We demonstrate that our model, which incorporates both multiplicative and additive sources of data dispersion, provides a plausible representation of realistic dispersion trends in regional methylation data. In addition, dSOMNiBUS simultaneously accounts for experimental errors, estimation of multiple covariate effects, and flexible dispersion patterns in a region. Also, we provide a formal inference for smooth covariate effects and construct a region-based statistic for the test of DMRs, where outcomes might be contaminated by errors and/or exhibit extra-parametric variations. Results from simulations and real data applications show that the new method captures important underlying methylation patterns with excellent power, provides accurate estimates of covariate effects, and correctly quantifies the underlying uncertainty in the estimates. The method has been implemented in the R package \texttt{SOMNiBUS}, which has been submitted to R Bioconductor.

Our model captures dispersion in the regional count data via the combination of a subject-specific RE and a multiplicative dispersion. The latter aims to capture the extra random dispersion beyond that introduced by the subject-to-subject variation. An alternative way to add multiplicative despersion might be to add locus-specific REs.  Such model would avoid the problem of estimating $\phi$, but would result in substantially increased number of REs, in which case our Laplace approximation is unlikely to provide well-founded inference \citep{shun1995laplace}. In addition, such a model only captures overdispersion.  In contrast, our quasi-binomial mixed effect model provides an adequate representation of any kind of dispersion without much increase in computational complexity.

An extension worth exploring in the future is to model the dispersion parameter $\phi$ as a function of covariates.  For example, the methylation variation across cancer samples has been found to be higher than for normal samples \citep{hansen2011increased,schoofs2013dna}. Identification of such disease-associated methylation variation changes might provide further insights into the biological mechanisms. This extension would also allow modelling of the hypothesis that some individuals are more sensitive to their environment \citep{meaney2005environmental}.
%SOME Michael Meaney's group hs written on this

Our proposed methods can also be applied to other types of next-generation sequencing data. For example, allele-specific gene expression (ASE) measured from RNA-seq data are quantified by the numbers of reads originating from the two alleles for that site \citep{ fan2020asep}. Such data share a similar structure to bisulfite sequencing data and could be analyzed by dSOMNiBUS. From the methodology point of view, our proposal of combining quasi likelihood with random effects can be generally applied to any type of count data for a more comprehensive representation of dispersion.

\section{Methods and Materials}
\label{method_detail}
%\subsection{Method details}
%\subsubsection{Estimation}
%\subsubsection{Inference for smooth covariate effects}
In this section, we present the methodology details on how to make inference about covariate effects $\beta_p(t_{ij})$ and simultaneously estimate the additive and multiplicative dispersion parameters $\phi$ and $\sigma_0^2$ in our smoothed quasi-binomial mixed model \eqref{model1}. %with outcomes measured with errors as defined in \eqref{error}.
We start with the case where true methylation counts $S_{ij}$ are available, and determine the complete data marginal quasi-likelihood function in Section \ref{complete_quasi-likelihood}. Then we describe the estimating algorithms for the complete and contaminated  data in Sections \ref{estimation} and \ref{estimation_with_error}, respectively. We additional estimate the pointwise CIs for covariate effects $\beta_p(t_{ij})$ and obtain tests of hypotheses for these effects in Section \ref{inference}.

%and proceed through more advanced techniques to obtain be  show the complete data quasi-likelihood in  Section \ref{complete_quasi-likelihood} and estimating algorithm   in Section \ref{estimation}, respectively.

\subsection{Laplace-approximated marginal quasi-likelihood function}
\label{complete_quasi-likelihood}
%We first reparameterize our model using basis representation and introduce the smoothness penalty, and then define the conditional, joint and marginal quasi-likelihood functions for the complete data $\left\{ \bs{S}, \bs{X}, \bs{Z}\right\}$.
\subsubsection{Basis representation}

In model \eqref{model1}, the function parameters $\beta_p(t_{ij})$ can be represented by the coefficients of chosen spline bases of rank $L_p$, 
$\beta_p(t_{ij}) = \sum_{l=1}^{L_p} \alpha_{pl} B_l^{(p)}(t_{ij}), \text{  for }p = 0, 1, \ldots P.
$ Here $ \left\{ B_l^{(p)} (\cdot)\right\}_{l=1}^{L_p}$ denotes the spline basis, and $\bs{\alpha_{p}} = (\alpha_{p1}, \ldots \alpha_{pL_p})^T \in \mathcal{R}^{L_p}$  are the coefficients to be estimated.
In this way, we can write the conditional mean in \eqref{model1} in a compact way as
	$$
	g^{-1}(\bs{\pi}) = \mathbb{X}^{(B)} \bs{\alpha} + \mathbb{X} ^{(1)} \bs{u},
	$$
	where $\bs{\pi} = (\pi_{11}, \ldots \pi_{1m_1},\pi_{21}, \ldots \pi_{2m_2}, \ldots \pi_{Nm_N})^{T} \in [0, 1]^M$ 
	%\in \mathcal{R}^{M}
	%is in a unit hypercube of dimension $M$, 
	%a $M$ hypercube of side 1, %SEARCH FOR THIS,
	with $M = \sum_{i=1}^N m_i$, $\bs{\alpha} = (\bs{\alpha_0, \alpha_1}, \ldots \bs{\alpha_p})^T \in \mathcal{R}^{K}$ with $K = \sum_{p=0}^{P}L_p$, and $\bs{u} = \left(  u_1, u_2, \ldots u_N\right)^T$.
	$\mathbb{X}^{{(B)}}$ is the spanned design matrix for $\bs{\alpha}$ of dimension $M \times K$, stacked with elements $ B_l^{(p)}(t_{ij}) \times Z_{pi}$ with $Z_{0i}\equiv 0$. $\mathbb{X}^{(1)}$ is a random effect model matrix of dimension $M \times N$,  with element 1 if the corresponding CpG site in the row belongs to the sample in the column, and 0 otherwise. If we write the overall spanned design matrix $\mathbb{X} = \left[\mathbb{X}^{(B)} , \;  \mathbb{X}^{(1)} \right] \in \mathcal{R}^{M \times (K+N)}$ and $\bs{\mathcal{B}} = (\bs{\alpha}^T, \bs{u}^T)^T$, the conditional mean can be further simplified as
		\begin{equation*}
		\label{neatmean}
		g^{-1}(\bs{\pi}) = \mathbb{X}\bs{\mathcal{B}}.
		\end{equation*}
	\subsubsection{Smoothness penalty}
To impose the assumption that the true covariate effect function is more likely to be smooth than jumpy, we add a smoothness penalty for each $\beta_p(t)$, $p = 0, 1, \ldots P$. The total amount of such penalty is an aggregate from all smooth terms, i.e.
	\begin{equation}
	\label{restriction}
	\mathcal{L}^{\rm{Smooth}} =  \sum_{p=0}^{P}  \lambda_p \int \left( \beta_p^{\prime \prime}\left(t\right)\right)^2 dt =
\sum_{p=0}^P \lambda_p \bs{\alpha_{p}}^{T}  \bs{A_p}  \bs{\alpha_p}
= \bs{\alpha^T A_\lambda \alpha},
%	=
%	 \dfrac{1}{\phi} \sum_{p=0}^P \lambda_p^{(s)} \bs{\alpha_{p}}^{T}  \bs{A_p}  \bs{\alpha_p} =  \dfrac{1}{\phi} \bs{\alpha}^T \bs{A}_\lambda \bs{\alpha}.
	\end{equation}
where $\bs{A_p}'s$ are $L_p \times L_p$ positive semidefinite matrices with the $(l, l^{\prime})$ element
$
\bs{A_p}{(l, l^\prime)} = \int {B^{(p)}}^{\prime \prime }_l(t) {B^{(p)}}^{\prime \prime }_{l^\prime}(t)dt,
$ which are fixed quantities given the specified set of bases.
The weights $\lambda_p$, i.e. the smoothing parameters, are positive parameters which establish a tradeoff between the closeness of the curve to the data and the smoothness of the fitted curves. $\bs{A}_\lambda$ is a $K \times K$ positive semidefinite block diagonal matrix of the form  $\bs{A_\lambda} = \text{Diag}\left\{ \lambda_0 \bs{A_0}, \lambda_1 \bs{A_1} , \ldots,  \lambda_P \bs{A_P} \right\}$.
%Here we introduce the scaling by $\phi$ is merely for later convenience.
\paragraph{Random-effect view of the smoothness penalty.}
As justified in \cite{wahba1983bayesian} and \cite{silverman1985some}, employing such smoothing penalty \eqref{restriction} during fitting is equivalent to imposing random effects for spline coefficients $\bs{\alpha}$. Specifically, $\bs{\alpha}$ is assumed to follow a (degenerate) multivariate normal distribution with precision matrix $\bs{A_\lambda}$,
\[
\bs{\alpha} \sim MVN(\bs{0}, \bs{A_\lambda}^{-}),
\]
where $\bs{A_\lambda}^{-}$ is the pseudoinverse of $\bs{A_\lambda}$. From a Bayesian viewpoint, imposing smoothness is equivalent to specifying a prior distribution on function roughness. This random-effect formulation of the smooth curve estimation problem opens up the possibility of estimating $\bs{\lambda}$ and $\phi$ using marginal (quasi-)likelihood maximization. In addition, under such a formulation, it requires no extra effort to estimate the `actual' RE term $\bs{u}$ in our model \eqref{model1}, once the inference procedure for $\bs{\alpha}$ is well established. In the rest of inference steps, we treat $\bs{\alpha}$ as random effects.
\subsubsection{Conditional quasi-likelihood function}
We first consider specifying the conditional ``distribution'' of $\bs{S}$ given the values of REs $\bs{\mathcal{B}}$. Following the notion of extended quasi-likelihood \citep[Section 9.6]{mccullagh1989generalized}, we define the following  conditional quasi-likelihood
\begin{equation}
\label{quasil}
qL^{\rm{(\bs{S}\mid\bs{\mathcal{B}})}}(\bs{\mathcal{B}}, \phi) \propto \exp\left\{ -\dfrac{1}{2\phi} \sum_{i,j}d_{ij}\left(S_{ij}, \pi_{ij} \right) -\dfrac{M}{2}\log{\phi} \right\},
\end{equation}
where 
\[
d_{ij}(S_{ij}, \pi_{ij}) = -2 \bigintssss_{S_{ij}/X_{ij}}^{\pi_{ij}} \dfrac{S_{ij}-X_{ij}\pi_{ij}}{\pi_{ij}(1-\pi_{ij})}d\pi_{ij}
\] is the quasi-deviance function corresponding to a single observation. 
%\[\
%\mathbb{E}\left\{ \dfrac{\partial\log\left[qL^{\rm{(\bs{S}\mid\bs{\gamma})}}(\bs{\gamma}, \phi)  \right]}{\partial \phi}\right\} \approx 0, 
%\] 
It can be easily checked that this quasi-likelihood exhibits the properties of log-likelihood, with respect to $\bs{\mathcal{B}}$. Such properties approximately hold for the dispersion parameter $\phi$, provided that $\phi$ be small and $\kappa_r = O(\phi^{r-1})$, where $\kappa_r$ is the rth-order cumulant of  $\bs{S}\mid\bs{B}$ \citep{efron1986double, jorgensen1987exponential, mccullagh1989generalized}.
Let $ql^{{(\bs{S}\mid\bs{\mathcal{B}})}}(\bs{\mathcal{B}}, \phi) = \log\left[ qL^{\rm{(\bs{S}\mid\bs{\mathcal{B}})}}(\bs{\mathcal{B}}, \phi)\right]$ denote the conditional log-quasi-likelihood. It should be noted that the integral inside $ql^{{(\bs{S}\mid \bs{\mathcal{B}})}}(\bs{\mathcal{B}}, \phi)$ rarely needs to be evaluated for the estimation of $\bs{\mathcal{B}}$, because the inference described later only requires the computation of its first and second derivatives, i.e.
 \[
 \dfrac{ \partial ql^{{(\bs{S}\mid\bs{\mathcal{B}})}}(\bs{\mathcal{B}}, \phi) }{\partial \bs{\mathcal{B}}} = \dfrac{1}{\phi} \mathbb{X}^T\left(\bs{S} - \bs{\Lambda_X}\bs{\pi}\right),
 \]
 \[
  \dfrac{\partial^2 ql^{{(\bs{S}\mid\bs{\mathcal{B}})}}(\bs{\mathcal{B}}, \phi)}{\partial \bs{\mathcal{B}} \partial\bs{\mathcal{B}}^T } =- \dfrac{1}{\phi} \mathbb{X}^T \bs{W} \mathbb{X},
 \]
 where $\bs{\Lambda_X} \in \mathbb{R}^{M\times M}$ is the diagonal matrix with values of read-depths, and $\bs{W}$ is the weight matrix whose diagonal is $X_{ij}\pi_{ij}(1-\pi_{ij})$.

\subsubsection{Joint quasi-likelihood functions}
For notational simplicity, we write $\bs{\Theta} = (\bs{\lambda}, {\sigma_0^2})$ for the parameters involved in the covariance structure of random effects $\bs{\mathcal{B}}$.
% Write $\bs{d} = (d_{11}, \ldots d_{1m_1},d_{21}, \ldots d_{2m_2}, \ldots d_{Nm_N})^{T} \in {\mathbb{R}}^M$
Combining the conditional `distribution' $\bs{S}\mid\bs{\mathcal{B}}$ with the marginal distribution of $\bs{\mathcal{B}}$, 
we obtain the following \textit{joint} log-quasi-likelihood of the observed data $\bs{S}$ and unobserved random effects $ \bs{\mathcal{B}}$ 
 \begin{eqnarray}
 \label{jointlik}
  q\ell^{(\bs{S}, \bs{\mathcal{B}})}(\bs{\mathcal{B}}, \phi, \bs{\Theta}) &=& ql^{(\bs{S}\mid \bs{\mathcal{B}})}(\bs{\mathcal{B}}, \phi) \underbrace{
    -\dfrac{1}{2} \bs{\alpha}^T \bs{A_\lambda} \bs{\alpha} -\dfrac{1}{2\sigma_0^2} \bs{u}^T\bs{u}}_{-\frac{1}{2\phi}\bs{\mathcal{B}}^T \bs{\Sigma_\Theta}\bs{\mathcal{B}}} \notag\\
  && + \underbrace{\dfrac{1}{2} \log \left\{  \lvert \bs{A_\lambda}\rvert_+\right\} + \dfrac{N}{2}\log\left(1/\sigma_0^2\right)}_{1/2\log \left\{  \lvert \bs{\Sigma_\Theta/\phi}\rvert_+\right\} },
 \end{eqnarray}
where $\bs{\Sigma_\Theta} = \text{diag}\left\{ \phi \bs{A_\lambda}, \phi/\sigma_0^2\bs{I}_N \right\}\in \mathbb{R}^{(K+N)\times(K+N)}$, and $ \lvert \bullet \rvert_+$ denotes the generalized determinant of a matrix, i.e. the product of its non-zero eigenvalues.
Here we introduce the scaling by $\phi$ in  $\bs{\Sigma_\Theta}$ merely for later convenience, and this allows us to factor out the dispersion parameter $\phi$ in the penalized quasi-score in \eqref{quasi_score}. In such way, the point estimates of random effects $\bs{\mathcal{B}}$ are independent of the estimate of $\phi$.
%The variance component parameters  thus become $\bs{\Theta^\star} = (\phi\bs{\lambda}, \sigma_0^2/\phi)$

This joint log-quasi-likelihood is composed of three parts: 1) the outcome `distribution' depending on $\bs{\mathcal{B}}$ and $\phi$, 2) multiple quadratic penalties for $\bs{\mathcal{B}}$ depending on regularization parameters $\bs{\Theta}$, and 3) fixed regularized terms for $\bs{\Theta}$. 
Our goals are to estimate the variance component parameters $\bs{\Theta}$, the dispersion parameter $\phi$, and also predict the values of random effects $\bs{\mathcal{B}}$. When $\phi = 1$, this fits a generalized linear mixed model (GLMM).
\begin{comment}
\paragraph{First derivative w.r.t. $\bs{\mathcal{B}}$}
\[
\dfrac{ \partial q\ell^{(\bs{S}, \bs{\mathcal{B}})}(\bs{\mathcal{B}}, \phi, \bs{\Theta}) }{\partial \bs{\mathcal{B}}} = \dfrac{1}{\phi} \mathbb{X}^T\left(\bs{S} - \bs{\Lambda_X}\bs{\pi}\right) - \bs{\Sigma_\Theta\mathcal{B}}
\]
\paragraph{Second derivative w.r.t. $\bs{\mathcal{B}}$}
\[
\dfrac{\partial^2 q\ell^{(\bs{S}, \bs{\mathcal{B}})}(\bs{\mathcal{B}}, \phi, \bs{\Theta})}{\partial \bs{\mathcal{B}} \partial\bs{\mathcal{B}}^T } = -\dfrac{1}{\phi} \mathbb{X}^T \bs{\Lambda_X} \mathbb{X} - \bs{\Sigma_\Theta}
\]
\end{comment}
\subsubsection{Laplace-approximated marginal quasi-likelihood function}
\label{laplace}
A legitimate (quasi-)likelihood is the \textit{marginal} `density' evaluated at the observed data $\bs{S}$ only, which is obtained by integrating out random effects $\bs{\mathcal{B}}$ from the joint quasi-likelihood of $\bs{S}$ and $\bs{\mathcal{B}}$,
\begin{equation}
\label{mlik}
qL^{M}(\phi, \bs{\Theta}) =  \bigintssss\exp \left\{
q\ell^{{(\bs{S}, \bs{\mathcal{B}})}}(\bs{\mathcal{B}}, \phi, \bs{\Theta}) 
\right\} d\bs{\mathcal{B}}.
\end{equation}
Conceptually, maximizing $qL^M(\phi, \bs{\Theta})$ yields the maximum quasi-likelihood estimators for $\bs{\Theta}$, and $\phi$. However, the analytical solutions for this high-dimensional integral are not easy to find, and an approximation approach is needed. 

%PROBABLY find more references using Laplace approimation to make inference for mixed-effect model
As in \cite{wood2011fast}, we use the Laplace approximation to evaluate the integral inside the marginal quasi-likelihood.
Let $\widehat{\bs{\mathcal{B}}}_{{\bs{\Theta}}}$ be the value of $\bs{\mathcal{B}}$ maximizing the joint quasi-likelihood $q\ell^{(\bs{S}, \bs{\mathcal{B}})}(\bs{\mathcal{B}}, \phi, \bs{\Theta})$ given the values of variance component parameters $\bs{\Theta}$, i.e.
\begin{equation}
\label{penlik}
\widehat{\bs{\mathcal{B}}}_{{\bs{\Theta}}} = \argmax\left\{ ql^{(\bs{S}\mid \bs{\mathcal{B}})}(\bs{\mathcal{B}}, \phi) {-\dfrac{1}{2\phi} \bs{\mathcal{B}}^T \bs{\Sigma_\Theta}\bs{\mathcal{B}}}  \right\},
\end{equation}
where terms not dependent on ${\bs{\mathcal{B}}}$ have been dropped from the joint quasi-likelihood. The objective function in \eqref{penlik} is often referred to as the penalized (quasi-)likelihood. A second-order Taylor expansion of  $q\ell^{(\bs{S}, \bs{\mathcal{B}})}(\bs{\mathcal{B}}, \phi, \bs{\Theta})$, around $\widehat{\bs{\mathcal{B}}}$ (the subscript $\bs{\Theta}$ has been dropped for notational simplicity), gives
\[
q\ell^{(\bs{S}, \bs{\mathcal{B}})}(\bs{\mathcal{B}}, \phi, \bs{\Theta}) \approx q\ell^{(\bs{S}, {\bs{\mathcal{B}}})}(\widehat{\bs{\mathcal{B}}}, \phi, \bs{\Theta}) - \dfrac{1}{2}\left(\bs{\mathcal{B}}-\widehat{\bs{\mathcal{B}}}\right)^T\bs{H_{\widehat{\mathcal{B}}}}\left(\bs{\mathcal{B}}-\widehat{\bs{\mathcal{B}}}\right),
\]
where $\bs{H_{\widehat{\mathcal{B}}}} = - \nabla^2_{\bs{\mathcal{B}}} \; {q}\ell^{(\bs{S}, \bs{\mathcal{B}})}(\widehat{\bs{\mathcal{B}}}, \phi, \bs{\Theta}) =  \dfrac{1}{\phi} \left(  \mathbb{X}^T \widehat{\bs{W}} \mathbb{X} + \bs{\Sigma_\Theta}\right)$.
Therefore, the \textit{marginal} quasi-likelihood in \eqref{mlik} can be approximately written as
\begin{eqnarray}
	qL^{M}(\phi, \bs{\Theta}) &\approx& \exp \left\{q\ell^{(\bs{S}, {\bs{\mathcal{B}}})}(\widehat{\bs{\mathcal{B}}}, \phi, \bs{\Theta})\right\}  \bigintssss\exp \left\{
	- \dfrac{1}{2}\left(\bs{\mathcal{B}}-\widehat{\bs{\mathcal{B}}}\right)^T\bs{H_{\widehat{\mathcal{B}}}}\left(\bs{\mathcal{B}}-\widehat{\bs{\mathcal{B}}}\right) 
	\right\} d\bs{\mathcal{B}} \notag\\
	& \approx&  \exp \left\{q\ell^{(\bs{S}, {\bs{\mathcal{B}}})}(\widehat{\bs{\mathcal{B}}}, \phi, \bs{\Theta})\right\}\dfrac{\sqrt{2\pi}^{K+N}}{\bigg\lvert  \dfrac{ \mathbb{X}^T \widehat{\bs{W}} \mathbb{X} + \bs{\Sigma_\Theta}}{\phi}\bigg\rvert^{1/2}} \notag \\
%	&\propto&qL^{\rm{(\bs{S}\mid\bs{\mathcal{B}})}}(\widehat{\bs{\mathcal{B}}}, \phi)\exp\left(-\dfrac{1}{2}\widehat{\bs{\mathcal{B}}}^T \bs{\Sigma_\Theta}\widehat{\bs{\mathcal{B}}}\right)\lvert   \bs{\Sigma_\Theta}\rvert^{1/2} \bigg\lvert  \dfrac{1}{\phi} \mathbb{X}^T \widehat{\bs{W}} \mathbb{X} + \bs{\Sigma_\Theta}\bigg\rvert^{-1/2} \notag \\
	\label{lapmar}
	&\propto& \phi^{-M/2} \exp\left(-\dfrac{\sum_{i,j}\widehat{d}_{ij}}{2\phi}\right)
	\exp\left(-\dfrac{1}{2\phi}\widehat{\bs{\mathcal{B}}}^T \bs{\Sigma_\Theta}\widehat{\bs{\mathcal{B}}}\right)\lvert   \bs{\Sigma_\Theta/\phi}\rvert_+                                   ^{1/2} \bigg\lvert  \dfrac{\mathbb{X}^T \widehat{\bs{W}} \mathbb{X} + \bs{\Sigma_\Theta}}{\phi} \bigg\rvert^{-1/2}. \notag \\
%	&\coloneqq& qL^{\rm{Laplace}}(\phi, \bs{\Theta}; \widehat{\bs{\mathcal{B}}})
\end{eqnarray}
In equation \eqref{lapmar}, $\widehat{d}_{ij} =d_{ij}(S_{ij}, \widehat{\pi}_{ij})$, where $\widehat{\pi}_{ij} = g^{-1}( \mathbb{X}_{(l,)}\widehat{\bs{\mathcal{B}}})$ and $l$ is the row in the model matrix $\mathbb{X}$ corresponding to CpG $j$ for sample $i$.
We denote this Laplace-approximated marginal quasi-likelihood in \eqref{lapmar} as $qL^{\rm{Laplace}}(\phi, \bs{\Theta}; \widehat{\bs{\mathcal{B}}})$ and simply write $\text{Laplace}(\phi, \bs{\Theta}; \widehat{\bs{\mathcal{B}}})$ $= \log[qL^{\rm{Laplace}}(\phi, \bs{\Theta}; \widehat{\bs{\mathcal{B}}})]$, which depends on $\bs{\Theta}$ via the dependence of $\bs{\Sigma_\Theta}$ and  $\widehat{\bs{\mathcal{B}}}$ (and thus $\widehat{\bs{W}}$ and $\widehat{\bs{d}}$) on $\bs{\Theta}$.
\subsection{Estimation algorithm for the complete data}
\label{estimation}
The essence of estimating $\bs{\Theta}, \bs{\mathcal{B}}$, and  $\phi$, is to optimize the Laplace-approximated marginal quasi-likelihood in \eqref{lapmar}. Note that such approximation requires calculating the maximum of the penalized quasi-likelihood in \eqref{penlik}, $\widehat{\bs{\mathcal{B}}}$, along with its corresponding Hessian $\bs{H_{\widehat{\mathcal{B}}}}$, which is only feasible for given values of the penalty parameters $\bs{\Theta}$. To disentangle the complicated dependence of $\widehat{\bs{\mathcal{B}}}$ on $\bs{\Theta}$, we adopt a nested-optimization strategy proposed by \cite{wood2011fast}. Specifically, the algorithm has an outer iteration for updating $\bs{\Theta}$ and $\phi$, with each iterative step supplementing with an inner iteration to estimate random effects $\bs{\mathcal{B}}$ corresponding to the current $\bs{\Theta}$, as summarized in Algorithm \ref{alg:no-error}. This Section proceeds with the detailed description of each step in Algorithm \ref{alg:no-error}.
\begin{algorithm}[h!]
	\SetKwData{Left}{left}\SetKwData{This}{this}\SetKwData{Up}{up}
	\SetKwFunction{Union}{Union}\SetKwFunction{FindCompress}{FindCompress}
	\SetKwInOut{Input}{Input}\SetKwInOut{Output}{Output}
	
	Initialize $\boldsymbol{\bs{\Theta}}^{(0)}, \phi^{(0)}$ ; Choose $\varepsilon=10^{-6}$; Set $s=0$\;
	
	\Repeat{$\|\boldsymbol{\bs{\mathcal{B}}}^{(s)}-\boldsymbol{\bs{\mathcal{B}}}^{(s-1)}\|_{2} < \varepsilon$}{
		Step 1. Solve $ \bs{U}(\bs{\mathcal{B}} ; \bs{\Theta}^{(s)}) =\bs{0}$ \eqref{quasi_score} to obtain $\bs{\mathcal{B}}^{(s)}$ \;
		%as defined in \eqref{quasi_score}
		Step 2. Newton's update for the Laplace-approximated marginal likelihood
		\footnotesize{
			$(\log(\phi),\log(\bs{\Theta}))^{(s+1)} = (\log(\phi),\log(\bs{\Theta}))^{(s)} - \left[ \nabla^2 \text{Laplace}({\bs{\mathcal{B}}}^{(s)}) \right]^{-1}\nabla\text{Laplace}({\bs{\mathcal{B}}}^{(s)})$}\;
		\normalsize{$s \leftarrow s+1$}\;
	}
	Return{\ $\boldsymbol{\Theta}^{(s)}, \bs{\mathcal{B}}^{(s)}, \phi^{(s)}$} \;  
	Step 3: Calculate $\widehat{\phi}_{Fle}$ using $\bs{\mathcal{B}}^{(s)}$
	
	\caption{Algorithm to find $(\widehat{\bs{\mathcal{B}}}, \widehat{{\phi}}, \widehat{\bs{\Theta}}) =  \argmax_{\bs{\mathcal{B}}, \phi, \bs{\Theta}}  \ell^{(\bs{S, \mathcal{B}})}\left( \bs{\mathcal{B}}, \phi, \bs{\Theta} \right)$ using data $\{\bs{S}, \bs{Z}, \bs{Y}\}$ \label{alg:no-error}}
\end{algorithm}
%with each iterative step 

\subsubsection{Inner iteration: estimate $\mathcal{B}$ given the current $\Theta$}
\label{inner}
Given the estimates of penalty parameters $\bs{\Theta}$, $\widehat{\bs{\mathcal{B}}}$ can be computed as the solution to 
\begin{equation}
\label{quasi_score}
\bs{U}\left(\bs{\mathcal{B}}\right) = \dfrac{1}{\phi}  \left\{ \mathbb{X}^T\left(\bs{S} - \bs{\Lambda_X}\bs{\pi}\right) - \bs{\Sigma_\Theta} \bs{\mathcal{B}}\right\}=\bs{0},
\end{equation}
where  $\bs{U}\left(\bs{\mathcal{B}}\right)$ is the \textit{quasi-score} for the penalized quasi-likelihood in \eqref{penlik} with respect to $\bs{\mathcal{B}}$. We use the Newton's method to solve these system of nonlinear equations. Specifically we compute the gradient of $\bs{U}\left(\bs{\mathcal{B}}\right)$,
\begin{equation*}
\label{obs_fisher}
\nabla \bs{U}\left(\bs{\mathcal{B}}\right) = - \dfrac{ \mathbb{X}^T {\bs{W}} \mathbb{X} + \bs{\Sigma_\Theta}}{\phi},
\end{equation*}
 and a single update from step $l$ to step $l+1$ for $\bs{\mathcal{B}}$ thus takes the form
\[
\bs{\mathcal{B}}^{(l+1)} = \bs{\mathcal{B}}^{(l)} + \left( \mathbb{X}^T {\bs{W}} \mathbb{X} + \bs{\Sigma_\Theta}\right)^{-1}\left[\mathbb{X}^T\left(\bs{S} - \bs{\Lambda_X}\bs{\pi}^{(l)}\right) - \bs{\Sigma_\Theta} \bs{\mathcal{B}}^{(l)}  \right].
\] 
%This Newton's iterative formula is analytically equivalent to a penalized  iteratively reweighted least squares (P-IRLS) update \citep[Section 3.4.1]{wood2017generalized}. 
We then iteratively update $\bs{\mathcal{B}}$ until convergence, which constitutes iteration Step 1 in Algorithm \ref{alg:no-error}.
\subsubsection{Outer iteration: maximize the Laplace-approximated marginal quasi-likelihood}
The outer iteration, which aims to maximize the Laplace-approximated marginal quasi-likelihood in \eqref{lapmar}, is also achieved by a Newton's method.  \cite{wood2011fast} has derived the derivatives and Hessian of $\text{Laplace}(\phi, \bs{\Theta}; \widehat{\bs{\mathcal{B}}})$  with respect to  $\bs{\rho} = (\log(\bs{\Theta}), \log(\phi))$, using a mixture of implicit and direct differentiations. We denote these first and second derivatives as $\nabla \text{Laplace}(\bs{\rho}; \widehat{\bs{\mathcal{B}}})$ and $\nabla^2 \text{Laplace}(\bs{\rho}; \widehat{\bs{\mathcal{B}}})$, respectively. Relying on the work of \cite{wood2011fast}, the maximization in the outer iteration can be readily achieved via
\begin{equation}
\label{step2}
\bs{\rho}^{(s+1)} = \bs{\rho}^{(s)}- \left[\nabla^2 \text{Laplace}\left(\bs{\rho}^{(s)};\widehat{\bs{\mathcal{B}}}^{(s)}\right)\right]^{-1}\nabla \text{Laplace}\left(\bs{\rho}^{(s)};\widehat{\bs{\mathcal{B}}}^{(s)}\right).
\end{equation}
Here, $\widehat{\bs{\mathcal{B}}}^{(s)}$ are the estimated mean parameters given the current $\bs{\Theta}^{(s)}$, obtained from the inner iteration in Section \ref{inner}. Each update in \eqref{step2} constitutes iteration Step 2 in Algorithm \ref{alg:no-error}.
We iterate between the Step 1 and Step 2 until convergence to obtain 
$\widehat{\bs{\mathcal{B}}}$, $\widehat{\bs{\Theta}}$ and $\widehat{\phi}$. 

%THE idea is in the case of GLMM, we can not factor out phi in the first derivative such that, lambda or theta can be estimated treating phi equal 1, the likelihood-based dispersion estimator (phi_REML) is important no matter which phi is used - expect the simulation results for folder "Exp_phi_3_re_9_quasi_RE_newb0_quasiF" is WORSE Sep 21 Actually the results is a bit conservative, with CI coverage almost 1

\subsubsection{Estimating $\phi$ using the moment-based estimator}
%\paragraph{Likelihood-based dispersion estimator} 
As described in the previous section, the dispersion parameter $\phi$ can be estimated as part of the outer iteration of the marginal quasi-likelihood maximization. We refer to this estimator as likelihood-based dispersion estimator, denoted as $\widehat{\phi}_{Lik}$.
%Estimate $\phi$ as part of the marginal quasi-likelihood maximization
%estimated by the default outer iteration, the scale estimator can be controlled using    
%\paragraph{Moment-based dispersion estimator}

In generalized linear models, it is common to estimate $\phi$ by dividing Pearson's lack-of-fit statistic by the residual degrees of freedom, and this is known as the moment-based scale/dispersion estimator.
We can apply the similar ideas here. Instead of using $\widehat{\phi}_{Lik}$, we take one step further and estimate $\phi$ using the final estimate $\widehat{\bs{\mathcal{B}}}$ (and thus $\widehat{\bs{\pi}}$). Specifically, Pearson's dispersion estimator can be written as
\[
\widehat{\phi}_P = \dfrac{1}{M-\tau} \sum_{i,j}\left(\dfrac{S_{ij}-X_{ij}\widehat{\pi}_{ij}}{\sqrt{X_{ij}\widehat{\pi}_{ij}(1-\widehat{\pi}_{ij})}} \right)^2.
\]
Here $\tau$ is the effective degrees of freedom \citep{wood2017generalized}, defined as
\begin{equation}
\label{traceF}
\tau = \text{trace}\left(\bs{F}\right), \; \text{with } \bs{F}=\left(\mathbb{X}^T\widehat{\bs{W}}\mathbb{X} + \bs{\Sigma_{\widehat{\Theta}}}\right)^{-1}\mathbb{X}^T\widehat{\bs{W}}\mathbb{X}.
\end{equation}
However, $\widehat{\phi}_P$ can be unstable at finite sample sizes, especially when a few Pearson residuals are huge \citep{farrington1995pearson, fletcher2012estimating}. For example, in our model, $\widehat{\pi}_{ij}$ close to $0$ can lead to a huge Pearson residual, even though the deviance $d_{ij}(S_{ij}, \widehat{\pi}_{ij})$ in \eqref{quasil} is modest.  Therefore, we adopt an improved version of the Pearson estimator, i.e. the Fletcher estimator \citep{fletcher2012estimating}, which is designed to mitigate this problem.  The Fletcher's dispersion estimator $\widehat{\phi}_{Fle}$ is defined as
\[
\widehat{\phi}_{Fle} = \dfrac{\widehat{\phi}_P}{1+\overline{a}}, \; \text{where } {a}_{ij} = \dfrac{1-2\widehat{\pi}_{ij}}{X_{ij}\widehat{\pi}_{ij}(1-\widehat{\pi}_{ij})}\left( S_{ij}-X_{ij}\widehat{\pi}_{ij}\right) \text{ and } \overline{a} = \dfrac{1}{M}\sum_{i,j}a_{ij}.
\]
If the mean model is adequate, then approximately we have 
\begin{equation}
\label{moment_chisq}
\dfrac{(M-\tau) \widehat{\phi}_{Fle}}{\phi} \sim \chi^2_{M-\tau}
\end{equation}
\citep{mccullagh1985asymptotic, fletcher2012estimating}. Therefore, $\widehat{\phi}_{Fle}$ provides an unbiased estimator for $\phi$, which is also confirmed by simulation results as shown in Supplementary Figure S9. In contrast, the estimation using $\widehat{\phi}_{Lik}$ can be considerably biased (Supplementary Figure S9). Hence, we calculate the moment-based estimate for the dispersion parameter, which constitutes the Step 3 in Algorithm \ref{alg:no-error}.
%ONE paragraph to justify why we prefer moment based estimator

\subsection{Estimating algorithm for the contaminated data}
\label{estimation_with_error}
In the presence of experimental errors, the true methylation data, $S_{ij}$ are unknown and one only observes $Y_{ij}$, which is assumed to be a mixture of binomial counts arising from both the truly methylated and truly unmethylated reads.
When $S_{ij}$ is modeled by a parametric distribution, like in \cite{zhao2020novel}, the EM algorithm \citep{dempster1977maximum} provides accurate estimation of the smooth covariate effects even though the true methylation data are missing. 
Motivated by the work of \cite{elashoff2004algorithm}, we propose an extension of the EM algorithm with special treatment for the multiplicative dispersion parameter $\phi$, to the case of quasi-likelihood-based analyses.

\subsubsection{Expectation-Solving algorithm}
%the transition sentence here is a bit awkward
\cite{elashoff2004algorithm} proposed an extension of the EM algorithm, called Expectation-Solving (ES) algorithm, to accommodate missing (or mis-measured) data when a natural set of estimating equations exists for the complete data setting. Specifically, the E step computes the conditional expectation of the estimating equations given the observed data, and S step solves these expected estimating equations.

To apply the ES algorithm to our case, we need to evaluate the conditional expectation of three sets of estimating equations:
\begin{eqnarray*}
\bs{U}(\bs{\mathcal{B}} ; \bs{\Theta}^{(s)}, \bs{S}) &=& \dfrac{1}{\phi}\left[\mathbb{X}^T\left(\bs{S} - \bs{\Lambda_X}\bs{\pi}^{(s)}\right) - \bs{\Sigma}_{\bs{\Theta}^{(s)}} \bs{\mathcal{B}}\right]  = \bs{0} \notag \\
 \nabla_{\bs{\Theta}}\text{Laplace}(\bs{\Theta}, \phi;{\bs{\mathcal{B}}}^{(s)},\bs{S}) 
 &=& \dfrac{1}{\phi}\sum_{i,j} \left\{  \dfrac{S_{ij}-X_{ij}\pi^{(s)}_{ij}}{\pi^{(s)}_{ij}(1-\pi^{(s)}_{ij})}\times \dfrac{d\pi^{(s)}_{ij}}{d\bs{\Theta}}\right\} + f_1(\bs{\Theta}, \phi;{\bs{\mathcal{B}}}^{(s)}) =  \bs{0} \notag \\
  \nabla_{\phi}\text{Laplace}(\bs{\Theta}, \phi;{\bs{\mathcal{B}}}^{(s)},\bs{S}) 
  &=& \dfrac{1}{\phi^2}\sum_{i,j}  \bigintssss_{S_{ij}/X_{ij}}^{\pi_{ij}^{(s)}} \dfrac{S_{ij}-X_{ij}\pi_{ij}}{\pi_{ij}(1-\pi_{ij})}d\pi_{ij} + f_2(\bs{\Theta}, \phi;{\bs{\mathcal{B}}}^{(s)}) =  \bs{0}, \notag
\end{eqnarray*}
for $\bs{\mathcal{B}}, \bs{\Theta}$ and $\phi$, respectively. Here, $\bs{\Theta}^{(s)}$, ${\bs{\mathcal{B}}}^{(s)}$, and $\bs{\pi}^{(s)}$ are estimates from the previous iterations, $f_1(\cdot)$ and $f_2(\cdot)$ denote the components that are independent of $\bs{S}$.

\paragraph{E step for $\bs{\mathcal{B}}$ and $\bs{S}$.}
The estimating equations for $\bs{\mathcal{B}}$ and $\bs{\Theta}$ are linear in the latent methylated counts $\bs{S}$, and thus their expectations equal $\bs{U}(\bs{\mathcal{B}} ; \bs{\Theta}^{(s)}, \bs{\eta}^\star)$ and  $\nabla_{\bs{\Theta}}\text{Laplace}(\bs{\Theta}, \phi;{\bs{\mathcal{B}}}^{(s)},\bs{\eta}^\star)$, respectively. Here, $\bs{\eta}^\star \in \mathcal{R}^M$ are the conditional expectations of $\bs{S}$ given $\bs{Y}$ evaluated at the trial estimates $\left( \bs{\mathcal{B}}^\star, \bs{\Theta}^\star \right)$, and for our model, take the form
\begin{comment}
\begin{equation*}
\label{estep}
\bs{\eta}^{\star} = \mathbb{E}\left( \bs{S} \mid {\bs{Y}} ; \bs{\mathcal{B}}^{\star}, \bs{\Theta}^\star \right)
=  \dfrac{ \bs{Y} p_1 \bs{\pi}^{\star} }{ p_1 {\bs{\pi}^{\star}} + p_0 (1-\bs{\pi}^{\star})} + \dfrac{\left(\bs{X}- \bs{Y}\right) (1-p_1)\bs{\pi}^{\star} }{ (1-p_1) \bs{\pi}^{\star} + (1-p_0) (1-\bs{\pi}^{\star})},
\end{equation*}
\end{comment}
\begin{equation}
\label{estep}
\eta_{ij}^{\star} = \mathbb{E}\left( S_{ij} \mid {Y_{ij}} ; \bs{\mathcal{B}}^{\star}, \bs{\Theta}^\star \right)
=  \dfrac{ Y_{ij} p_1 \pi_{ij}^{\star} }{ p_1 {\pi_{ij}^{\star}} + p_0 (1-\pi_{ij}^{\star})} + \dfrac{\left(X_{ij}- Y_{ij}\right) (1-p_1)\pi_{ij}^{\star} }{ (1-p_1) \pi_{ij}^{\star} + (1-p_0) (1-\pi_{ij}^{\star})},
\end{equation}
where ${\pi}^\star_{ij} = g^{-1}( \mathbb{X}_{(l,)}{\bs{\mathcal{B}}^\star})$ and $l$ is the row in the model matrix $\mathbb{X}$ corresponding to CpG $j$ for sample $i$. These expected estimating equations can then be solved using the direct nested iteration method in Algorithm \ref{alg:no-error}.
%Iterating between the E and S steps provides $\widehat{\bs{\mathcal{B}}}$ and $\widehat{\bs{\Theta}}$ for the contaminated data
\paragraph{E step for $\phi$.}
However, the estimating equation for $\phi$ is not linear in the unknown methylated counts $\bs{S}$; see details in Appendix \ref{non_linear}. Therefore, the closed-form exact expression for $\mathbb{E}_{\bs{S}\mid{Y}; \bs{\mathcal{B}}^\star,\bs{\Theta}^\star}(\nabla_{\phi}\text{Laplace}(\bs{\Theta}, \phi;{\bs{\mathcal{B}}}^{(s)},\bs{S}))$ is not available, and the E-S algorithm cannot be readily applied to estimating $\phi$ from the contaminated data. To circumvent this problem, we propose a direct method to estimate $\phi$ without undergoing the E-S iteration. 
%Then $\bs{\mathcal{B}}$ and $\bs{\Theta}$ are estimated by iterating between the E and S steps treating $\phi$ as known.
\subsubsection{A plug-in estimator for $\phi$}
Specifically, we estimate $\phi$ by exploiting its relationship with the dispersion for the observed outcome $\bs{Y}$, denoted as $\phi^Y_{ij}$, which is defined as
\[
\phi^Y_{ij} = \dfrac{\mathbb{V}\text{ar}(Y_{ij} \mid u_i)}{X_{ij}{\pi^Y_{ij}} (1-{\pi^Y_{ij}})}, \text{ with } {\pi^Y_{ij}} = \mathbb{E}(Y_{ij}\mid u_i) = \pi_{ij}p_1 + (1-\pi_{ij})p_0.
\]
Based on our assumed mean-variance relationship \eqref{conditiona_var} and error model \eqref{error}, we can express $\phi^Y_{ij}$ in terms of $\phi$, $\pi_{ij}$ and error parameters $p_0$ and $p_1$,
\begin{equation}
\label{phi_relation}
\phi^Y_{ij} = 1+ (\phi - 1) \dfrac{(\pi^Y_{ij}-p_0)(p_1-\pi^Y_{ij})}{{\pi^Y_{ij}} (1-{\pi^Y_{ij}})}; 
\end{equation}
see detailed derivations in Appendix \ref{derivphi}. Although we assume a constant dispersion $\phi$ for the true outcome $\bs{S}$, the observed outcome $\bs{Y}$ implied by our error model, possesses dispersion parameter $\phi^Y_{ij}$ varying with each CpG site, when $\phi \neq 1$.

Directly running the nested iteration method (Algorithm \ref{alg:no-error}) on the observed data $\{\bs{Y, Z, X}\}$ reports a constant dispersion estimate $\widehat{\phi}^Y$ and $\widehat{\pi}^Y_{ij}$ for all $i$ and $j$, along with other useful estimates. We assume that $\widehat{\phi}^Y$ is an estimate for the mean of individual dispersions $\phi^Y_{ij}$, i.e. 
\begin{equation}
\label{phi_relation_mean}
\frac{1}{M}\sum_{i,j}\phi^Y_{ij} = 1 + (\phi -1)\dfrac{1}{M} \sum_{i,j}\dfrac{({\pi}^Y_{ij}-p_0)(p_1-{\pi}^Y_{ij})}{{{\pi}^Y_{ij}} (1-{{\pi}^Y_{ij}})};
\end{equation}
empirical results show that this is a reasonable assumption, as shown in Supplementary Figure S11.
We then propose to estimate $\phi$ by plugging in the error-prone outcome-related estimates $\widehat{\phi}^Y$ and $\widehat{\pi}^Y_{ij}$ to the relation in \eqref{phi_relation_mean}:
%$ = 1 + (\phi-1) \dfrac{1}{M} \sum_{i,j}\dfrac{(\pi^Y_{ij}-p_0)(p_1-\pi^Y_{ij})}{{\pi^Y_{ij}} (1-{\pi^Y_{ij}})}$. 
\[
\widehat{\phi} = (\widehat{\phi}^Y -1)\left[\dfrac{1}{M} \sum_{i,j}\dfrac{(\widehat{\pi}^Y_{ij}-p_0)(p_1-\widehat{\pi}^Y_{ij})}{{\widehat{\pi}^Y_{ij}} (1-{\widehat{\pi}^Y_{ij}})} \right]^{-1} + 1.
\]
%The plug-in approach relates the asymptotic properties of 
%COMMNET on this estimator
\subsubsection{A hybrid ES algorithm}
We propose a hybrid ES algorithm to estimate our model using the error-prone outcomes $\bs{Y}$. We first estimate $\phi$ using the aforementioned plug-in approach and then estimate $\bs{\mathcal{B}}$ and $\bs{\Theta}$ using ES iterations assuming $\phi$ is fixed and known; detailed steps are summarized in Algorithm \ref{alg:with-error}. We denote the final estimates from our algorithm as $\widehat{\phi}$, $\widehat{\bs{\mathcal{B}}}$ and $\widehat{\bs{\Theta}}$.
The components of $\widehat{\bs{\alpha}}$ inside the vector of $\widehat{\bs{\mathcal{B}}}$ leads to estimates of the functional parameters $\beta_p(t)$, for $p = 0, 1, \ldots, P$:
\[
\widehat{\beta_p(t)} = \left\{\bs{B}^{(p)}(t)\right\}^{T} \left\{ \widehat{\bs{\alpha}_p} \right\}, 
\]
where $t$ is a genomic position lying within the range of the input positions $\left\{ t_{ij} \right\}$, and $\bs{B}^{(p)}(t) = ( B_1^{(p)}(t), B_2^{(p)}(t), \ldots B_{L_p}^{(p)}(t))^T \in \mathcal{R}^{L_p}$ is a column vector with nonrandom quantities obtained from evaluating the set of basis functions $ \{ B_l^{(p)} (\cdot)\}_l$ at position $t$.
%These expected estimating equations can then be solved using the direct nested iteration method in Algorithm \ref{alg:no-error} 

%Iterating between the E and S steps provides $\widehat{\bs{\mathcal{B}}}$ and $\widehat{\bs{\Theta}}$ for the contaminated data

%Calculating these conditional expectations $\eta_{ij}^\star$ from \eqref{estep} constitutes the E step of our algorithm.
\begin{algorithm}[h!]
	\SetKwData{Left}{left}\SetKwData{This}{this}\SetKwData{Up}{up}
	\SetKwFunction{Union}{Union}\SetKwFunction{FindCompress}{FindCompress}
	\SetKwInOut{Input}{Input}\SetKwInOut{Output}{Output}
	Step 1: run Algorithm \ref{alg:no-error} on $\{\bs{Y, Z, X}\}$; return $\bs{\widehat{\pi}}^Y$, $\widehat{\phi}_Y$, $\bs{\widehat{B}}$, and $\widehat{\bs{\Theta}}$\;
	Step 2: calculate the plug-in estimator $\widehat{\phi}$ \; 
	Step 3: E-S iterations with $\phi$ fixed at $\widehat{\phi}$ to estimate $\bs{\mathcal{B}}$ and $\bs{\Theta}$; specifically \ \ 
Initialize $\boldsymbol{\bs{\Theta}}^{(0)} = \widehat{\bs{\Theta}}, , \bs{\mathcal{B}}^{(0)} = \bs{\widehat{B}}$; Choose $\varepsilon=10^{-6}$; Set $\ell=0$\;
	\Repeat{$\|\boldsymbol{\bs{\mathcal{B}}}^{(\ell)}-\boldsymbol{\bs{\mathcal{B}}}^{(\ell-1)}\|_{2} < \varepsilon$}{
		$\bullet$	E step: $\eta_{ij}^{(\ell)} = \mathbb{E}(S_{ij} \mid Y_{ij}; \bs{\mathcal{B}}^{(\ell)})$\;
		$\bullet$ S step: $({\bs{\mathcal{B}}^{(\ell)}},  {\bs{\Theta}^{(\ell)}}) =  \argmax_{\bs{\mathcal{B}}, \bs{\Theta}}  \ell^{(\bs{\mathcal{B}}, \bs{\Theta})}\left( \bs{\mathcal{B}},  \bs{\Theta} ; \eta_{ij}^{(\ell)},\widehat{\phi} \right)$. Specifically\qquad 
		\Repeat{$\|\boldsymbol{\bs{\mathcal{B}}}^{(s)}-\boldsymbol{\bs{\mathcal{B}}}^{(s-1)}\|_{2} < \varepsilon$}{
			$\bullet$	Solve $ \bs{U}(\bs{\mathcal{B}} ; \bs{\Theta}^{(s)};\bs{\eta}^{(\ell)}) =\bs{0}$ to obtain $\bs{\mathcal{B}}^{(s)}$ using data $\eta_{ij}^{(\ell)}$ \;
			%as defined in \eqref{quasi_score}
			$\bullet$	Newton's update for the Laplace approximated marginal likelihood evaluated at data $\eta_{ij}^{(\ell)}$ :
			$(\log\bs{\Theta})^{(s+1)} = (\log\bs{\Theta})^{(s)} - \left[ \nabla^2_{\bs{\Theta}} \text{Laplace}({\bs{\mathcal{B}}}^{(s)}) \right]^{-1}\nabla_{\bs{\Theta}}\text{Laplace}({\bs{\mathcal{B}}}^{(s)})$\;
			$s \leftarrow s+1$\;
		}
		$\ell \leftarrow \ell+1$\;	
	}
	Return{\ $\boldsymbol{\Theta}^{(\ell)}, \bs{\mathcal{B}}^{(\ell)}$}  \;
	
	\caption{A hybrid ES algorithm to estimate the smoothed quasi-binomial mixed model with error-prone outcomes. \label{alg:with-error}}
\end{algorithm}

\begin{comment}
 1) $\bs{U}(\bs{\mathcal{B}} ; \bs{\Theta}^{(s)}, \bs{S}) =\bs{0}$ for $\bs{\mathcal{B}}$, where $\bs{U}$ is the quasi score as defined in \eqref{quasi_score},  2)
 \[
 \nabla_{\bs{\Theta}}\text{Laplace}(\bs{\Theta}, \phi;{\bs{\mathcal{B}}}^{(s)},\bs{S}) = \dfrac{1}{\phi}\sum_{i,j} \left\{  \dfrac{S_{ij}-X_{ij}\pi^{(s)}_{ij}}{\pi^{(s)}_{ij}(1-\pi^{(s)}_{ij})}\times \dfrac{d\pi^{(s)}_{ij}}{d\bs{\Theta}}\right\} + f(\bs{\Theta}, \phi;{\bs{\mathcal{B}}}^{(s)}) =  \bs{0}
 \] for $\bs{\Theta}$, where $f(\bs{\Theta}, \phi;{\bs{\mathcal{B}}}^{(s)})$ denotes the rest of parts that are independent of $\bs{S}$ and 3)
 \[
 \nabla_{\phi}\text{Laplace}(\bs{\Theta}, \phi;{\bs{\mathcal{B}}}^{(s)}, \bs{S}) =\bs{0}
 \] for $\phi$. Here $\bs{\Theta}^{(s)}$,and ${\bs{\mathcal{B}}}^{(s)}$ are estimates from the previous iterations, and we additional include
\end{comment}

\begin{comment}
When the true outcomes $\bs{S}$ are available, one can maximize the \textit{complete} joint likelihood in \eqref{jointlik} to obtain estimates for $\bs{\mathcal{B}}, \phi,$ and $\bs{\Theta}$, using the nested iteration optimization presented in  Algorithm \ref{alg:no-error}. Specifically, Algorithm \ref{alg:no-error} alternates between the steps of solving $\bs{U}(\bs{\mathcal{B}} ; \bs{\Theta}^{(s)}, \bs{S}) =\bs{0}$ for $\bs{\mathcal{B}}$ and solving $\nabla\text{Laplace}(\bs{\Theta}, \phi;{\bs{\mathcal{B}}}^{(s)}, \bs{S}) =\bs{0}$ for $\bs{\Theta}$ and $\phi$, where $\bs{\Theta}^{(s)}$ and ${\bs{\mathcal{B}}}^{(s)}$ are estimates from the previous iterations.
\end{comment}

\subsection{Inference for smooth covariate effects}
\label{inference}
We then estimate the pointwise confidence intervals (CI) for the smoothed covariate effects $\left\{\beta_1(t), \beta_2(t), \ldots, \beta_P(t)\right\}$, and obtain tests of hypotheses for these effects. Note that the inference is carried out conditional on the values of variance component parameters $\bs{\Theta}$ and dispersion parameter $\phi$, i.e. the uncertainty in estimating them is not accounted for. 
\subsubsection{Estimating the variance of the resulting parameter estimates}
As did in \cite{elashoff2004algorithm}, we can re-express the E step as the solution to the following M-dimensional estimating equation:
\begin{eqnarray*}
	\bs{U}^{(2)} (\bs{S}) &=& \bs{S} - \widehat{\bs{\eta}} = \bs{0},
\end{eqnarray*}
where $\widehat{\bs{\eta}}$ are the conditional expectations in \eqref{estep} evaluated at the current estimate $\widehat{\bs{\pi}}$.
In this way, the overall ES algorithm can be viewed as solving an expanded set of equations of dimension $K+N+M$, whose first $K+N$ components are $\bs{U}\left(\bs{\mathcal{B}}\right)=\bs{0}$ in \eqref{quasi_score} and whose second $M$ components are $\bs{U}^{(2)} (\bs{S}) =\bs{0}$.

Under this formulation, we use the established theory for estimating equations \citep{lindsay1982conditional, heyde1996quasi, small2003numerical}, and propose a model-based variance estimator for $\widehat{\bs{\mathcal{B}}}$. Specifically, under correct specification of the first two moments of $\bs{S}$, the asymptotic variance of $\widehat{\bs{\mathcal{B}}}$ can be written as
\[
{\mathbb{V}\text{ar}}(\widehat{\bs{\mathcal{B}}}) =  \left[ (-\bs{D})^{-1}\right]_{(\bs{\mathcal{B}}, \bs{\mathcal{B}})},
\]
where $\bs{D}$ is the first order derivative of the expanded estimating equations for $\bs{\mathcal{B}}$ and $\bs{S}$, and $[\; \bullet \;]_{(\bs{\mathcal{B}}, \bs{\mathcal{B}})}$ stands for the matrix block corresponding to $\bs{\mathcal{B}}$. In our case, $\bs{D}$ takes the form
\[
\bs{D}  = -  \left[
\begin{matrix}
\dfrac{1}{\phi} \mathbb{X}^T \bs{W} \mathbb{X} + \dfrac{1}{\phi} \bs{\Sigma_\Theta}  &&  -\dfrac{1}{\phi}  \mathbb{X} ^T \\
\bs{W_\delta}  \mathbb{X}  && - \bs{I}_{M}.
\end{matrix}
\right] 
 \]
Here, $\bs{W_\delta}$ is a diagonal matrix with elements $X_{ij}\delta_{ij}$, where
\begin{equation*}
	\label{delta}
	\delta_{ij} = 
	\dfrac{ Y_{ij} p_1 p_0 }{ \left[  p_1 {\pi_{ij}^{}} + p_0 (1-\pi_{ij}^{}) \right]^2} + \dfrac{\left(X_{ij}- Y_{ij}\right) (1-p_1)(1-p_0) }{ \left[  (1-p_1) \pi_{ij}^{} + (1-p_0) (1-\pi_{ij}^{}) \right] ^2},
\end{equation*}
and reduces to a zero matrix when $p_0 = 1-p_1 = 0$. 
Then, the asymptotic variance of $\widehat{\bs{\mathcal{B}}}$ can be simplified as
\begin{equation}
\label{asyvarfor}
{\mathbb{V}\text{ar}}(\widehat{\bs{\mathcal{B}}}) =  \left[\mathbb{X}^T{({\bs{W}}-\bs{W_\delta})}\mathbb{X}+\bs{\Sigma_{{\Theta}}}\right]^{-1} {\phi}.
\end{equation}
Therefore, the desired variance estimator of $\widehat{\bs{\mathcal{B}}}$ can be obtained by plugging in the final estimates $\widehat{\bs{\mathcal{B}}}, \widehat{\bs{\Theta}}$ and $\widehat{\phi}$ into equation \eqref{asyvarfor}. 

\begin{comment}

\citep{small2003numerical}

In quasi-likelihood inference, correct specification of the first two moments of the underlying distribution leads to the information unbiasedness, which states that two forms of the information matrix: the negative sensitivity matrix (negative expectation of the first order derivative of an estimating function) and the variability matrix (variance of an estimating estimating function) are equal. 

Model-based covariance matrix estimator and sandwich covariance matrix estimator,
\end{comment}

\subsubsection{Confidence interval estimation}
\label{CI}
Let $\bs{\widehat{V}}$ denote the aforementioned variance estimator and $\bs{\widehat{V}_p}$ be the diagonal blocks of $\widehat{\bs{V}}$ corresponding to $\bs{\alpha}_p$, with dimensions $L_p \times L_p$. We then immediately have  the estimated variance of $\widehat{\beta_p(t)}$:
$
\widehat{\mathbb{V}\text{ar}} (\widehat{\beta_p(t)})  =  \left\{\bs{B}^{(p)}(t)\right\}^{T}  \bs{\widehat{V}_p}  \left\{\bs{B}^{(p)}(t)\right\}.
$
Therefore, the confidence interval for $\beta_p(t)$ at significance level $\nu$ can be approximately estimated by
$
\widehat{\beta_p(t)} \pm \mathbb{Z}_{\nu/2} \sqrt{\widehat{\mathbb{V}\text{ar}} (\widehat{\beta_p(t)}) }, 
$
for any $t$ in the range of interest, where $\mathbb{Z}_{\nu/2}$ is $\nu/2$ (upper-tail) quantile of a standard normal distribution.
\subsubsection{Hypothesis testing for a regional zero effect}
 \label{test}
 We can also construct a region-wide test of the null hypothesis 
 \begin{equation*}
 H_0: \beta_p(t) = {0}, \text{for any } t \text{ in the genomic interval}.
 \end{equation*}
 This test depends on the association between covariate $Z_p$ and methylation levels across the region, after adjustment for all the other covariates, and the null hypothesis is equivalent to  $H_0: \bs{\alpha}_p = \bs{0}$. We propose the following region-based F statistic
  \begin{equation*}
  \label{F}
  T_p =  \dfrac{\widehat{\bs{\alpha_p}}^T \left\{ \bs{\widehat{V}_p}  \right\}	^{-1} \widehat{\bs{\alpha_p}}}{\tau_p},
  \end{equation*}
 where $\{ \bs{\widehat{V}_p}\}^{-1}$ denotes inverse if $\bs{\widehat{V}_p} $ is nonsigular; for singular  $\bs{\widehat{V}}_p$, the inverse is replaced by the Moore-Penrose inverse $\{ \bs{\widehat{V}_p} \}^{-}$. Here, $\tau_p$ is the effective degrees of freedom (EDF) for smooth term $\beta_p(t)$, which depends on the magnitude of smoothing parameter $\bs{\lambda}$ and random effect variances $\sigma_0^2$.
  Motivated by the work of \cite{wood2013p}, we define the EDF $\tau_p$ as 
  \begin{equation}
  \label{edf}
  \tau_p  =   \sum_{l=a_p }^{b_p} \left(2\bs{F}-\bs{F} \bs{F}  \right)_{(l,l)}, \text{ for } p = 0, 1, \ldots P, \notag
  \end{equation}
  where  $a_p =\sum_{m=0}^{p-1}L_m+1$ if $p >0$ and $a_p =1$ if $p =0 $, $ b_p = \sum_{m=0}^p L_m$ for any $p$, and $\left( \bullet \right)_{(l,l)}$ stands for the $l^{th}$ leading diagonal element of a matrix.  $\bs{F}$ is the smoothing matrix of our model, as defined in \eqref{traceF}, which can be viewed as the matrix mapping the pseudo data to its predicted mean.

Let $\bs{V}_p =\widehat{\bs{V}}_p \cdot \phi/ \widehat{\phi} $ be the variance estimator for $\bs{\alpha}_p$ when the dispersion parameter $\phi$ is known. \cite{zhao2020novel} have shown the following asymptotic results under the null
\[
	\widehat{\bs{\alpha_p}}^T \left\{ \bs{{V}_p}  \right\}	^{-1} \widehat{\bs{\alpha_p}} \sim \chi^2_{\tau_p}.
\]
Combining with the property of moment-based dispersion estimator in \eqref{moment_chisq}, we can conclude that, under the null hypothesis, $T_p$ asymptotically follows a F distribution with degrees of freedom $\tau_p$ and $M-\tau$, i.e. 
$T_p \sim F_{\tau_p, M-\tau}$.
%Therefore, a well-calibrated null distribution for the Wald statistic $T_p$ is a Chi-square distribution with degree of freedom equal to $\tau_p$.

\bibliography{mybib}
\bibliographystyle{apalike}

\appendix
\section{Marginal interpretations for dSOMNiBUS}

\subsection{Marginal mean}
\label{deriv_mean}
The latent variable representation of the logistic mixed effect model in \eqref{model1} is
\begin{eqnarray*}
S_{ijk}^\star &=& \eta_{ij} + \epsilon_{ij} + u_i \\
S_{ijk} &=& \begin{cases}
	1, & \text{ if }  S_{ijk}^\star  \geq 0\\
	0, &  \text{ if }  S_{ijk}^\star  < 0
\end{cases} 
\end{eqnarray*}
where $S_{ijk}^\star$ is the unobserved latent variable, $\eta_{ij} = \sum_{p=0}^P \beta_p(t_{ij})Z_{pi}$ is the linear predictor calculated from all the fixed effect, $\epsilon_{ij}$ are iid error terms following a logistic distribution, and $u_i$ is the subject-specific random effect as defined in Section \ref{model-section}. In addition, the error term $\epsilon_{ij}$ and RE $u_i$ are mutually independent. Specifically,  the cumulative distribution function (cdf) for $\epsilon$ takes the form $g(x) = 1/(1+\exp(-x))$.  The calculation of marginal mean $\pi_{ij}^M = \mathbb{P}(\eta_{ij} + \epsilon_{ij} + u_i \geq 0)$ requires integration over the joint distribution of  $\epsilon_{ij}$ and $u_i$, which has no closed-form solution. Instead, we can approximate the logistic cdf $g(x)$ by a normal cdf \citep[p. 119]{johnson1995continuous}, which will lead to a more analytically tractable solution. Specifically, we have
\[
g(x) \approx \Phi(cx), \text{ with } c = \sqrt{3.41}/\pi,
\]
where $\Phi(x)$ is the cdf of the standard  normal distribution. For any $x$ value, the maximum absolute difference of this approximation is 0.00948.

Therefore, we can approximately view $\epsilon_{ij}$ as a normal random variable, $\epsilon_{ij} \sim N(0, 1/c^2)$. Since $\epsilon_{ij}$ and $u_i$ are independent, we have $\epsilon_{ij} + u_i \sim N(0, 1/c^2 + \sigma_0^2)$. The marginal mean can be thus derived as
\begin{eqnarray*}
	\pi_{ij}^M &=& \mathbb{P}(\epsilon_{ij} +u_i \geq - \eta_{ij})  = \mathbb{P}\left( \dfrac{\epsilon_{ij} + u_i}{\sqrt{1/c^2+\sigma_0^2}}  \geq \dfrac{-\eta_{ij}}{\sqrt{1/c^2+\sigma_0^2}} \right) \\
	&\approx& \Phi \left( \dfrac{\eta_{ij}}{\sqrt{1/c^2+\sigma_0^2}} \right) \approx g\left(  \dfrac{\eta_{ij}}{\sqrt{1+c^2\sigma_0^2}}. \right)
\end{eqnarray*}

\subsection{Marginal variance}
\label{deriv_var}
We will use the mixed effect model formulation in \eqref{model1} to derive the marginal variance. Using the law of total variance, the marginal variance of $S_{ij}$ is the sum of two parts:
\begin{eqnarray}
\label{total_law}
\mathbb{V}ar(S_{ij}) &=& \mathbb{E}\left\{ \mathbb{V}ar(S_{ij} \mid u_i) \right\} + \mathbb{V}ar\left\{ \mathbb{E}(S_{ij} \mid u_i) \right\}  \notag \\
&=& \phi X_{ij}\mathbb{E} \left\{ \pi_{ij} (1-\pi_{ij})  \right\} + X_{ij}^2 \mathbb{V}ar \left( \pi_{ij} \right), 
\end{eqnarray}
where $\pi_{ij} = g(\eta_{ij} + u_i )$ is the conditional mean dependent on $u_i$.  The exact closed-form formula does not exist for either $\mathbb{E}\left( \pi_{ij}\right)$ or $\mathbb{V}ar \left( \pi_{ij} \right)$. Nevertheless, 
%as did in \cite{goldstein1991nonlinear}, 
we can work on the second-order Taylor expansion of $\pi_{ij}$ around $u_i =0$, i.e. 
$\pi_{ij} =  g\left( \eta_{ij}+ u_i \right)
\approx g\left(  \eta_{ij} \right) +  g^{\prime} \left(  \eta_{ij} \right) u_i + g^{\prime \prime} \left(  \eta_{ij} \right) u_i^2/2 $. Thus, we have $\mathbb{E}(\pi_{ij}) \approx g(\eta_{ij}) + g^{\prime \prime} \left(  \eta_{ij} \right) \sigma_0^2/2 $,
\begin{eqnarray*}
\mathbb{V}ar \left( \pi_{ij} \right) &\approx& \mathbb{E}\left\{ \left[
g^{\prime} (  \eta_{ij} ) u_i + \dfrac{g^{\prime \prime}(\eta_{ij})}{2} \left( u_i^2 - \sigma_0^2 \right) \right]^2 \right\}\\
&=&  \sigma_0^2 \left[  g^{\prime} (  \eta_{ij} )\right]^2 + \dfrac{\sigma_0^4}{2} \left[  g^{\prime\prime} (  \eta_{ij} )\right]^2,
\end{eqnarray*}
and  $\mathbb{E}\left( \pi_{ij}^2 \right) \approx \sigma_0^2 \left[  g^{\prime} (  \eta_{ij} )\right]^2 + \dfrac{\sigma_0^4}{2} \left[  g^{\prime\prime} (  \eta_{ij} )\right]^2 + \left[g(\eta_{ij})+ \dfrac{g^{\prime \prime} t(  \eta_{ij} )}{2} \sigma_0^2 \right]^2 $. Substituting the above approximations into \eqref{total_law} yields the results in equation \eqref{marginal_var}.

%, the exact analytical solution for such marginal model is not available with the logit link. Nevertheless, as did in \cite{goldstein1991nonlinear}, we can derive the first-order approximation to the marginal mean and variance of $S_{ij}$ implied by the mixed effect model. 

%A key feature of the mixed effect model in \eqref{model1} is that the smooth covariate effect $\beta_p(t_{ij})$  needs to be interpreted conditionally on the value of random effect $u_i$, i.e. interpreting at the level of individual subjects. If one desires estimates of covariate effect on population average, it is required to derive the corresponding marginal model by integrating out the random effect $\bs{u}$. However, the exact analytical solution for such marginal model is not available with the logit link. Nevertheless, as did in \cite{goldstein1991nonlinear}, we can derive the first-order approximation to the marginal mean and variance of $S_{ij}$ implied by the mixed effect model. 

%Define $g(x) = logit^{-1}(x)$ and $\eta_i(t_{ij})=\sum_{p=0}^P \beta_p(t_{ij})Z_{pi}$, with $Z_{0i} \equiv 0$.
%We can approximately rewrite the model in \eqref{model1} as
%\[ 
%{S_{ij}}\approx X_{ij} g\left(  \eta_i+ u_i \right) +\epsilon_i, \;\; \text{where }  \mathbb{E}(\epsilon_i)=0 \text{ and }  \mathbb{V}\text{ar}(\epsilon_i)=\phi   X_{ij} \pi_{ij} (1-\pi_{ij}).
%\]
%We can then use a first-order Taylor expansion of $\pi_{ij} =g  \left(  \eta_i+ u_i  \right)$ around $u_i=0$ and have
%\[ 
%{S_{ij}} \mid u_i \approx X_{ij} \left\{  g(\eta_i) + g^\prime (\eta_i) u_i  \right\} +\epsilon_i, 
%\]
\section{Estimate $\phi$ from the contaminated data}

\subsection{No exact expression available for the E step for $\phi$}
\label{non_linear}
Once evaluated the integral in the quasi-deviance $d_{ij}(S_{ij}, {\pi}_{ij})$ \eqref{quasil}, the estimating equation for $\phi$ takes the form 
\begin{eqnarray*}
\nabla_{\phi}\text{Laplace}(\bs{\Theta}, \phi;{\bs{\mathcal{B}}}^{(s)},\bs{S}) 
&=& \dfrac{1}{\phi^2}\sum_{i,j}  \bigintssss_{S_{ij}/X_{ij}}^{\pi_{ij}^{(s)}} \dfrac{S_{ij}-X_{ij}\pi_{ij}}{\pi_{ij}(1-\pi_{ij})}d\pi_{ij} + f_2(\bs{\Theta}, \phi;{\bs{\mathcal{B}}}^{(s)})\\
&=& \dfrac{1}{\phi^2}\sum_{i,j} \left\{
 (X_{ij}-S_{ij})\log(1-\pi_{ij}^{(s)}) + S_{ij} \log(\pi_{ij}^{(s)}) \right. \\
 && \left.- (X_{ij}-S_{ij})\log(1-S_{ij}/X_{ij}) - S_{ij} \log(S_{ij}/X_{ij}^{})  \right\}+ f_2(\bs{\Theta}, \phi;{\bs{\mathcal{B}}}^{(s)}).
\end{eqnarray*}
This estimating equation is not linear in terms of the unknown methylated counts $\bs{S}$. Thus, replacing $S_{ij}$ by $\eta_{ij}^\star= \mathbb{E}\left( S_{ij} \mid {Y_{ij}} ; \bs{\mathcal{B}}^{\star}, \bs{\Theta}^\star \right)$ does not necessarily provide an accurate estimate for $\mathbb{E}_{\bs{S}\mid{Y};\bs{\Theta}^\star, \bs{\mathcal{B}}^\star}(\nabla_{\phi}\text{Laplace}(\bs{\Theta}, \phi;{\bs{\mathcal{B}}}^{(s)},\bs{S}))$, and the exact expression for this expectation is not readily available from the first two moments of the distribution of $S_{ij}$.

\subsection{The relation between $\phi^Y_{ij}$ and $\phi$}
\label{derivphi}
All the expectation and variance in this section are conditional on the values of random effects $u_i$. For notational simplicity, we drop $u_i$ from all the derivations in this section.

The variance of $Y_{ij}$ depends on its mean $\pi^Y_{ij}$ as well as the joint probability $\mathbb{P}(Y_{ijk}=1, Y_{ijk^\prime} = 1)$, i.e. observing methylated signals at both the $k^{th}$ and ${k^\prime}^{th}$ reads, where $k, k^\prime = 1, 2, \ldots X_{ij}$ and $k \neq k^\prime$:
\begin{eqnarray}
\mathbb{V}\text{ar}(Y_{ij}) &=& \mathbb{E}(Y_{ij}^2) - \left[\mathbb{E}(Y_{ij})\right]^2  =  \mathbb{E}\left\{\left(\sum_{k=1}^{X_{ij}} Y_{ijk}\right)^2\right\} - X_{ij}^2({\pi^Y_{ij}})^2 \notag \\
\label{vary}
&=& \sum_{k=1}^{X_{ij}} \mathbb{E}(Y_{ijk}^2) + 2 \sum_{k=1}^{X_{ij}} \sum_{k^\prime=1}^{k-1}\mathbb{E}(Y_{ijk}Y_{ijk^\prime}) - X_{ij}^2({\pi^Y_{ij}})^2 \notag \\
&=& X_{ij}{\pi^Y_{ij}} -X_{ij}^2 ({\pi^Y_{ij}})^2 +2 \sum_{k=1}^{X_{ij}} \sum_{k^\prime=1}^{k-1}\mathbb{P}(Y_{ijk}=1, Y_{ijk^\prime} = 1).
\end{eqnarray}
By the law of total probability, we have
\[
\mathbb{P}(Y_{ijk}= Y_{ijk^\prime} = 1)=\sum_{s_1=0}^1\sum_{s_2=0}^1 \mathbb{P}(S_{ijk}=s_1, S_{ijk^\prime}=s_2)\mathbb{P}(Y_{ijk} = Y_{ijk^\prime}=1 \mid S_{ijk}=s_1,  S_{ijk^\prime}=s_2).
\] 
%where we assume that given the true methylation states $S_{ijk}$ and $S_{ijk^\prime}$, the observed methylation states $Y_{ijk}$ and $Y_{ijk^\prime}$ are independent.
\paragraph{Joint distribution of the bivariate outcomes $(S_{ijk}, S_{ijk^\prime})$.}
Note that, under our assumed mean-variance relationship in \eqref{conditiona_var}, $S_{ijk}$ and $S_{ijk^\prime}$ are not necessarily independent. Define $a_{ijkk\prime} =\mathbb{P}(S_{ijk}=1, S_{ijk^\prime}=1)$. The joint probability mass function of $(S_{ijk}, S_{ijk^\prime})$ can be thus written as
\begin{eqnarray*}
	\mathbb{P}(S_{ijk}=1, S_{ijk^\prime}=1)  &=& a_{ijkk\prime}\\
	\mathbb{P}(S_{ijk}=1, S_{ijk^\prime}=0)  &=& \pi_{ij} - a_{ijkk\prime}\\
	\mathbb{P}(S_{ijk}=0, S_{ijk^\prime}=1)  &=& \pi_{ij}-a_{ijkk\prime}\\
	\mathbb{P}(S_{ijk}=0, S_{ijk^\prime}=0)  &=& 1- 2\pi_{ij}+a_{ijkk\prime}.
\end{eqnarray*}
We now can write the probability of observing two methylated reads as
\begin{equation}
\label{proby}
\mathbb{P}(Y_{ijk}=Y_{ijk^\prime} = 1)=p_0^2(1- 2\pi_{ij}+a_{ijkk\prime})+2p_0p_1(\pi_{ij} - a_{ijkk\prime})+p_1^2a_{ijkk\prime}. \notag
\end{equation}
Here, we assume that given the true methylation states $S_{ijk}$ and $S_{ijk^\prime}$, the observed methylation states $Y_{ijk}$ and $Y_{ijk^\prime}$ are independent.
\paragraph{Derive the values of $a_{ijkk\prime}$.}
From first principle, we can express the variance of $S_{ij}=\sum_{k=1}^{X_{ij}} S_{ijk}$,
\begin{eqnarray}
\mathbb{V}\text{ar}(S_{ij}) &=& \sum_{k=1}^{X_{ij}} \mathbb{V}\text{ar}(S_{ijk}) + 2 \sum_{k=1}^{X_{ij}} \sum_{k^\prime=1}^{k-1}\mathbb{C}\text{ov}(S_{ijk}, S_{ijk^\prime}) \notag \\
&=& X_{ij}\pi_{ij}(1-\pi_{ij}) +  2 \sum_{k=1}^{X_{ij}} \sum_{k^\prime=1}^{k-1}\mathbb{E}(S_{ijk} S_{ijk^\prime}) - 2 \sum_{k=1}^{X_{ij}} \sum_{k^\prime=1}^{k-1}\mathbb{E}(S_{ijk})\mathbb{E}( S_{ijk^\prime}) \notag \\
&=&  X_{ij}\pi_{ij}(1-\pi_{ij}) +  2 \sum_{k=1}^{X_{ij}} \sum_{k^\prime=1}^{k-1}\mathbb{P}(S_{ijk}=1, S_{ijk^\prime}=1)   - X_{ij}(X_{ij}-1)\pi_{ij}^2 \notag \\
\label{var2}
&=&  X_{ij}\pi_{ij}(1-\pi_{ij}) +   2 \sum_{k=1}^{X_{ij}} \sum_{k^\prime=1}^{k-1}a_{ijkk\prime}  - X_{ij}(X_{ij}-1)\pi_{ij}^2. \notag
\end{eqnarray}
On the other hand, we have $\mathbb{V}\text{ar}(S_{ij}) = \phi X_{ij}\pi_{ij}(1-\pi_{ij})$. Equating these two quantities gives 
\begin{equation}
\label{aij}
2 \sum_{k=1}^{X_{ij}} \sum_{k^\prime=1}^{k-1}a_{ijkk\prime} = 
 (\phi-1)X_{ij}\pi_{ij}(1-\pi_{ij})+X_{ij}(X_{ij}-1)\pi_{ij}^2 \notag
\end{equation}
\paragraph{Derive $\mathbb{V}\text{ar}(Y_{ij})$ and $\phi_Y$.}
Now, we can plug the expression of $\mathbb{P}(Y_{ijk}=Y_{ijk^\prime} = 1)$ in \eqref{vary} and write  $\mathbb{V}\text{ar}(Y_{ij})$ in terms of $\phi$
\begin{eqnarray*}
\mathbb{V}\text{ar}(Y_{ij}) &=& X_{ij}{\pi^Y_{ij}} - X_{ij}^2 ({\pi^Y_{ij}})^2 + 
2 \sum_{k=1}^{X_{ij}} \sum_{k^\prime=1}^{k-1}\left[p_0^2(1- 2\pi_{ij}+a_{ijkk\prime})+2p_0p_1(\pi_{ij} - a_{ijkk\prime})+p_1^2a_{ijkk\prime} \right]\\
&=& X_{ij}{\pi^Y_{ij}} - X_{ij}^2  ({\pi^Y_{ij}})^2 + X_{ij}(X_{ij}-1)\left\{p_0^2(1-2\pi_{ij})+2p_0p_1\pi_{ij}\right\} + 2 (p_0-p_1)^2 \sum_{k=1}^{X_{ij}} \sum_{k^\prime=1}^{k-1}a_{ijkk\prime}. \\
&=& X_{ij}{\pi^Y_{ij}} - X_{ij}^2  ({\pi^Y_{ij}})^2 + X_{ij}(X_{ij}-1)\left\{p_0^2(1-2\pi_{ij})+2p_0p_1\pi_{ij}\right\}\\
&& + (p_0-p_1)^2 \left\{ (\phi-1)X_{ij}\pi_{ij}(1-\pi_{ij})+X_{ij}(X_{ij}-1)\pi_{ij}^2\right\}\\
%&=&  X_{ij}{\pi^Y_{ij}} - X_{ij}^2  ({\pi^Y_{ij}})^2 +X_{ij}(X_{ij}-1)\left[ p_0(1-\pi_{ij}) + p_1\pi\right]^2 +  (p_0-p_1)^2(\phi-1)X_{ij}\pi_{ij}(1-\pi_{ij})\\
%&=&  X_{ij}{\pi^Y_{ij}} - X_{ij}^2  ({\pi^Y_{ij}})^2 +X_{ij}(X_{ij}-1)({\pi^Y_{ij}})^2 +  (p_0-p_1)^2(\phi-1)X_{ij}\pi_{ij}(1-\pi_{ij})\\
&=& X_{ij}{\pi^Y_{ij}}(1-{\pi^Y_{ij}}) +  (p_0-p_1)^2(\phi-1)X_{ij}\pi_{ij}(1-\pi_{ij})
\end{eqnarray*}
The multiplicative dispersion parameter for the mis-measured outcome $Y$ is thus
\[
\phi^Y_{ij} = \dfrac{\mathbb{V}\text{ar}(Y_{ij})}{X_{ij}{\pi^Y_{ij}}(1-{\pi^Y_{ij}})} = 1+ (\phi -1)\dfrac{\pi_{ij}(1-\pi_{ij})}{{\pi^Y_{ij}}(1-{\pi^Y_{ij}})}(p_0-p_1)^2.
\]
Plugging in $\pi_{ij} = \dfrac{\pi^Y_{ij}-p_0}{p_1-p_0}$ leads to the relation in \eqref{phi_relation}.    

\end{document}